%
%

\input harvmac
\newcount\yearltd\yearltd=\year\advance\yearltd by 0
\noblackbox

\input epsf
\def\tilde{\widetilde}
\def\hat{\widehat}
\newcount\figno\figno=0

\def\fig#1#2#3{
\par\begingroup\parindent=0pt\leftskip=1cm\rightskip=1cm\parindent=0pt
\baselineskip=11pt \global\advance\figno by 1 \midinsert
\epsfxsize=#3 \centerline{\epsfbox{#2}} \vskip 12pt {\bf Figure\
\the\figno: } #1\par
\endinsert\endgroup\par
}
\def\figlabel#1{\xdef#1{\the\figno}}
\def\encadremath#1{\vbox{\hrule\hbox{\vrule\kern8pt\vbox{\kern8pt\hbox{$\displaystyle #1$}\kern8pt}\kern8pt\vrule}\hrule}}

\def\half{{1\over 2}}

 \def\d{{\delta}}

 \def\a{{\alpha}}
 
 \def\frac#1#2{{#1\over #2}}
 \def\l{{\lambda}}

 \def\g{{\gamma}}
 \def\s{{\sigma}}
 
 \def\b{{\beta}}

\def\bal{{\bar{\alpha}}}

\def\bg{{\bar{\gamma}}}
\def\bd{{\bar{\delta}}}

\def\e{{\epsilon}}

\def\IR{\relax{\rm I\kern-.18em R}}
\def\tdzero{D_{\tau}}
\def\tdzero{D_{\tau}}
\def\ja{{j_{\alpha}}}
\def\jb{{j_{\beta}}}
\def\jc{{j_{\gamma}}}


\lref\sundborg{ B.~Sundborg, ``The Hagedorn transition,
deconfinement and $N = 4$ SYM theory,'' Nucl.\ Phys.\ B {\bf 573},
349 (2000) [arXiv:hep-th/9908001].
}

\lref\first{ O.~Aharony, J.~Marsano, S.~Minwalla, K.~Papadodimas and
M.~Van Raamsdonk, ``The Hagedorn / deconfinement phase transition in
weakly coupled large $N$ gauge theories,'' Adv.\ Theor.\ Math.\
Phys.\  {\bf 8}, 603 (2004)
  [arXiv:hep-th/0310285].
}

\lref\firstord{
O.~Aharony, J.~Marsano, S.~Minwalla, K.~Papadodimas and M.~Van Raamsdonk,
  ``A first order deconfinement transition in large N Yang-Mills theory on a
  small $S^3$,''
  Phys.\ Rev.\ D {\bf 71}, 125018 (2005)
  [arXiv:hep-th/0502149].
}

\lref\beisert{ N.~Beisert, G.~Ferretti, R.~Heise and K.~Zarembo,
``One-loop QCD spin chain and its spectrum,''
  Nucl.\ Phys.\ B {\bf 717}, 137 (2005)
  [arXiv:hep-th/0412029].
}

\lref\spradlin{ M.~Spradlin and A.~Volovich,
``A pendant for Polya: The one-loop partition function of $N = 4$ SYM on
$R \times S^3$,''
  Nucl.\ Phys.\ B {\bf 711}, 199 (2005)
  [arXiv:hep-th/0408178].
}

\lref\thooft{
 G.~'t Hooft,
  ``A planar diagram theory for strong interactions,''
  Nucl.\ Phys.\ B {\bf 72}, 461 (1974).
}

\lref\Cutkosky{ R.~E.~Cutkosky, ``Harmonic functions and matrix
elements for hyperspherical quantum field models,'' J.\ Math.\
Phys.\  {\bf 25}, 939 (1984).
}

\lref\svv{
  M.~Spradlin, M.~Van Raamsdonk and A.~Volovich,
  ``Two-Loop Partition Function In The Planar Plane-Wave Matrix Model,''
  Phys.\ Lett.\ B {\bf 603}, 239 (2004)
  [arXiv:hep-th/0409178].
}

\lref\splitone{G.~Leibbrandt and J.~Williams, 
``Split Dimensional Regularization for the Coulomb Gauge,''
Nucl.\ Phys.\ B {\bf 475}, 469 (1996)  [arXiv:hep-th/9601046].
}

\lref\splittwo{  Y.~H.~Chen, R.~J.~Hsieh and C.~l.~Lin,  
``Split dimensional regularization for the temporal gauge,''  
arXiv:hep-th/9610165.  
}

\lref\splitthree{G.~Leibbrandt,   
``The three-point function in split dimensional regularization in the  
Coulomb gauge,''  
Nucl.\ Phys.\ B {\bf 521}, 383 (1998)  
[arXiv:hep-th/9804109].  
}

\lref\splitfour{G.~Heinrich and G.~Leibbrandt,
``Split dimensional regularization for the Coulomb gauge at two loops,''  
Nucl.\ Phys.\ B {\bf 575}, 359 (2000)  
[arXiv:hep-th/9911211].  
}

\def\my_Title#1#2{\nopagenumbers\abstractfont\hsize=\hstitle\rightline{#1}%
\vskip .5in\centerline{\titlefont #2}\abstractfont\vskip .5in\pageno=0}

\my_Title {\vbox{\baselineskip12pt \hbox{WIS/10/06-AUG-DPP}
\hbox{\tt hep-th/0608156}}} {\vbox{\centerline{Two Loop Partition
Function for} \vskip 5pt \centerline{Large $N$ Pure Yang-Mills
Theory on a Small $S^3$}}}

\centerline{Ofer Aharony$^{a}$, Joseph Marsano$^{b}$, and Mark Van Raamsdonk$^{c}$}

\medskip

\centerline{\sl $^{a}$Department of Particle Physics, Weizmann Institute of
Science, Rehovot 76100, Israel}
\centerline{\sl $^{b}$Jefferson Physical Laboratory, Harvard University,
Cambridge, MA 02138, USA}
\centerline{\sl $^{c}$Department of Physics and Astronomy,
University of British Columbia,}
\centerline{\sl Vancouver, BC, V6T 1Z1, Canada}
\medskip


\medskip

\noindent We give a direct path-integral calculation of the
partition function for pure $3+1$ dimensional $U(N)$ Yang-Mills
theory at large $N$ on a small $S^3$, up to two-loop order in
perturbation theory. From this, we calculate the one-loop shift in
the Hagedorn/deconfinement temperature for the theory at small
volume, finding that it increases (in units of the inverse sphere
radius) as we go to larger coupling (larger volume). Our results
also allow us to read off the sum of one-loop anomalous dimensions
for all operators with a given engineering dimension in planar
Yang-Mills theory on $\IR^4$. As checks on our calculation, we
reproduce both the Hagedorn shift and some of the anomalous
dimension sums by independent methods using the results of {\tt
hep-th/0412029} and {\tt hep-th/0408178}. The success of our
calculation provides a significant check of methods used in {\tt
hep-th/0502149} to establish a first order deconfinement transition
for pure Yang-Mills theory on a small $S^3$.

\Date{}  

\newsec{Introduction}

Pure $3+1$ dimensional Yang-Mills theory on $S^3$ has a discrete
energy spectrum which, thanks to asymptotic freedom, can be computed
perturbatively for small volume. This information is encoded in the
thermal partition function, which can be used to evaluate
thermodynamic functions and investigate the phase structure of the
theory. In this note, we explicitly compute the partition function
at two-loop order in perturbation theory for the $U(N)$ Yang-Mills
theory in the 't Hooft large $N$ limit \thooft,
by a direct evaluation of the Euclidean path
integral on $S^3 \times S^1$, where the radius of $S^3$ is $R_{S^3} \ll
1/\Lambda_{QCD}$ and the circumference of the circle is the inverse
temperature $\beta$.

Our results for the partition function may be expanded in powers of
the dimensionless variable $x\equiv e^{-\beta / R_{S^3}}$ as
\eqn\result{ Z_{2 \; loop}(x) = \sum_{n=2}^{\infty} x^n (a_n +
\lambda b_n \ln(x)+ {\cal O}( \lambda^2)), }
where $a_n$ gives the number of states with energy $n/R_{S^3}$ in the free
theory
and $b_n \lambda$ gives the sum of the order $\lambda$ perturbative
energy corrections for all states with energy $n/R_{S^3}$ ($\lambda$ is
the 't Hooft coupling $g_{YM}^2 N$). Equivalently, $b_n \lambda$
gives the sum of one-loop anomalous dimensions for all dimension $n$
operators in the planar Euclidean Yang-Mills theory on $\IR^4$.
Using spin chain techniques, the matrix of one-loop anomalous
dimensions has previously been computed in \beisert, so we can in
principle derive the $b_n$'s from those results also. We find
agreement between the two methods for all cases that we have checked
explicitly.

As discussed in \refs{\sundborg,\first}, large $N$ $U(N)$ (or
$SU(N)$) Yang-Mills theory in the limit of small volume has a
Hagedorn density of states, manifested as a divergence of the
partition function at a critical temperature given by 
$T_{c,0} = [R_{S^3}\ln(2+\sqrt{3})]^{-1}$
at $\lambda=0$.  At large (but not infinite) $N$, this Hagedorn
divergence signals a deconfinement phase transition across which the
free energy jumps from order one to order $N^2$. As a physical
application of our results, we can evaluate the leading perturbative
correction to this critical temperature\foot{Note that beyond first
order in perturbation theory, we expect the phase transition
temperature to be below the Hagedorn temperature \firstord, but to
order $\lambda$, they are the same.} by determining the point at
which our two-loop partition function diverges. We find that it
increases (in units of the inverse radius) as the coupling increases
\foot{Since the coupling runs, we should clarify that the relevant
coupling is the one at the scale $1/R_{S^3}$, so an increase in the
coupling is the same as an increase in $R_{S^3}\Lambda_{QCD}$.}:
\eqn\hagtemp{T_c R_{S^3} = T_{c,0}R_{S^3} \cdot 
\left(1 + {\lambda \over {12 \pi^2}} + {\cal
O}(\lambda^2)\right) \; .}
This is consistent with the conjecture that $T_c R_{S^3}$ is monotonic in
the radius, since at large radius we expect $T_c$ to approach a
constant of order $\Lambda_{QCD}$, so $T_c R_{S^3}$ is an increasing
function of $R_{S^3}$. As another check of our results, we provide an
independent calculation of the shift in $T_c$ using a formula in
\spradlin\ for the shift in the Hagedorn temperature in terms of the
matrix of anomalous dimensions, which we can get from \beisert.
Again, we find that the two methods give the same result.

While our results are interesting in their own right, one of the
main motivations for this work was to provide a check of the
calculation and regularization methods used in \firstord\ to
establish that the planar Yang-Mills theory on a small $S^3$
undergoes a first-order deconfinement transition as the temperature
is increased. That calculation involved a number of novel features
related to the spherical space: sums over angular momenta instead of
spatial momentum integrals, integrals over three-vector or scalar
spherical harmonics on $S^3$ appearing in the vertices, and a novel
non-gauge-invariant cutoff regularization scheme used together with
counterterms chosen to restore gauge invariance in physical
calculations. While the calculation in \firstord\ had internal
consistency checks, such as the cancellation of all logarithmic
divergences among the diagrams, we did not have any independent
method of verifying the final numerical result (the sign of which
determined the order of the phase transition). Since the
calculational setup and many of the steps are identical in the
present calculation,\foot{The calculation in \firstord\ was more
specific in that the necessary information was contained in specific two and
three-loop contributions to the effective action for the constant
mode of $A_0$ on $S^3 \times S^1$ after the path integral was
performed over all other (massive) degrees of freedom.} we view the
successful matching of our results here with those from other
methods (involving only standard calculations on $\IR^4$) as a
satisfying check of the validity of our formalism\foot{In fact, the
computation described in this paper helped us to find a small
mistake in the original version of the calculation of \first. This
mistake did not affect the main result of \first.}.

The structure of the paper is as follows: in section 2, we provide
the basic setup for our calculation and outline the two-loop
diagrams that will be needed in order to compute the partition
function. These are divergent, so we need in addition certain
one-loop counterterm diagrams. In section 3, we review our
regularization scheme and calculate the necessary counterterms,
including a new curvature-dependent counterterm not present for
flat-space calculations. In section 4, we evaluate the regularized
two-loop diagrams together with the one-loop counterterm diagrams in
order to get our final result for the partition function and,
subsequently, for the one-loop correction to the Hagedorn
temperature. We verify that our results are independent of the
regulator used. In section 5, we verify our results by independent
calculations of the Hagedorn shift and of the first several terms in
the expansion \result\ using results from \spradlin\ and \beisert.

\newsec{Setup}

The basic setup for our calculation is identical to that in \firstord,
so we give only a brief summary here, referring the reader to section
2 of that paper and section 4 of \first\ for more details.

We would like to calculate the thermal partition function
\eqn\partfunc{ Z = \sum_i e^{- \beta E_i} = \sum_i e^{-\beta
\Delta_i / R_{S^3}} = \sum_i x^{\Delta_i}}
for pure $U(N)$ Yang-Mills theory on $S^3$ with radius $R_{S^3}$, where
$\beta$ is the inverse temperature, $\Delta_i$ are the dimensions of
the local operators in the theory, and $E_i$ are the energy levels.
This is given by the Euclidean path integral on $S^3 \times S^1$,
with the circle (compactified Euclidean time) chosen to have
circumference $\beta$, of the Euclidean Yang-Mills action
\eqn\seuc{S_{Euc} = {1\over 4} \int_0^{\beta} dt \int d^3x \,
\tr(F_{\mu \nu} F^{\mu \nu}) \; .}
We use a normalization in which the Yang-Mills coupling $g_{YM}$
appears in the interaction terms in the action. To make the path
integral well-defined, we need to fix a gauge and include the
appropriate Fadeev-Popov determinants. For our calculation on $S^3$,
it is convenient to use the gauge
\eqn\gf{ \del_i A^i=0, \qquad \qquad
\partial_t \int_{S^3} A_0=0 \; , }
where $i$ runs over the spatial $S^3$ directions.
The latter condition implies that
the zero-mode of $A_0$ on $S^3$ is constant in time.  We denote
this mode by $\alpha$.

Apart from $\alpha$, which has no quadratic term in the Yang-Mills
action \seuc, all other modes are massive.  As such, it is convenient to
first integrate them out to obtain an effective action for $\alpha$.
As argued in \first, this effective action can only depend on the
unitary matrix $U=e^{i \beta \alpha}$ (the Wilson line of the gauge
field around the thermal circle, averaged over the sphere). Further,
the effect of the Fadeev-Popov determinant associated with the
second gauge-fixing condition in \gf\ is to convert the measure $[d
\alpha]$ in the path integral to the Haar measure $[dU]$. Thus, we
obtain
\eqn\firstz{ Z = \int [dU] e^{-S_{eff}(U)}}
where
\eqn\secondz{e^{-S_{eff}(U)} = \int [dA'][dc][d\bar{c}]
e^{-S_{Euc}(A',\alpha) - S_{FP}(A,c)}.}
Here, $[dA']$ denotes the measure for the gauge fields excluding the
zero mode of $A_0$. The fields $c$ and $\bar{c}$ are ghosts and
\eqn\sfp{S_{FP} = - \int_0^\beta dt \int_{S^3} \tr(\del_i {\bar c}
{\hat D}^i c)}
is the ghost action (where ${\hat D}^i$ is a covariant derivative),
introduced to represent the Fadeev-Popov determinant associated with
the first gauge-fixing condition in \gf.

\subsec{Vertices and propagators}

To perform the calculation, we expand all fields into modes on $S^3$
(we set the radius of $S^3$ to be $R_{S^3}=1$ from here on as it can always
be reinstated by dimensional analysis),
\eqn\expand{ \eqalign{ A_0(t,\theta) & = \sum_{\alpha}
a^{\alpha}(t) S^{\alpha}(\theta); \cr A_i(t,\theta) &= \sum_{\beta}
A^{\beta}(t) V_i^{\beta}(\theta); \cr c(t,\theta) &= \sum_{\alpha}
c^{\alpha}(t) S^{\alpha}(\theta), \cr} }
where $S^\alpha$ and $V_i^\beta$ are scalar and vector spherical
harmonics on $S^3$, defined in appendix B ($\alpha$ and $\beta$
represent angular momentum quantum numbers). This leads to the
action in appendix A. Because the fields $a^\alpha$ and $c^\alpha$ appear
only quadratically, we can integrate them out first
to get an effective action involving only $A^\beta$ and $\alpha$.
The final result includes a quadratic action
\eqn\quadratic{ S_2 =
\int dt 
\, \tr\left( {1 \over 2} A^{\bar \alpha} ( - \tdzero^2 +
(j_\alpha+1)^2) A^\alpha \right), }
a cubic interaction
\eqn\cubic{S_3 = g_{YM} \int dt \, \tr(i
 A^\alpha A^\beta A^\gamma \epsilon_\alpha (j_\alpha + 1)
E^{\alpha \beta \gamma}),}
and quartic interactions
\eqn\quartic{\eqalign{ S_4 = g_{YM}^2 \int dt \, \tr & \left(- {1 \over
2} A^\alpha A^\beta A^\gamma A^\delta \left( D^{\alpha \gamma
\bar{\lambda}} D^{\beta \delta \lambda} - D^{\alpha \delta
\bar{\lambda}} D^{\beta \gamma \lambda} \right) \right. \cr & \quad\left. + {1 \over 2}
{D^{\alpha_1 \beta_1 \gamma} D^{\alpha_2 \beta_2 \bar{\gamma}} \over
j_\gamma (j_\gamma + 2)} [A^{\alpha_1}, D_\tau A^{\beta_1}]
[A^{\alpha_2}, D_\tau A^{\beta_2}]\right)\; ,}}
where we have defined $D_\tau \equiv \partial_t - i [\alpha,
\cdot]$, and $D^{\alpha \beta \gamma}$
and $E^{\alpha \beta \gamma}$ are integrals over products of three
spherical harmonics that are defined in appendix B. Sums over all indices are
implied.

There are in addition higher order vertices arising (like the second
line in \quartic) from integrating out $a$ and $c$, but these do not
enter in our two-loop calculation. We have also suppressed a set of
vertices proportional to $\delta(0)$ which serve to precisely cancel
terms proportional to $\delta(0)$ that arise in contracting two
$D_\tau A$'s at equal times. As explained in \firstord, we can
simply work with the set of interactions above, ignoring any
$\delta(0)$ terms that arise.\foot{These $\delta(0)$'s are related
to our choice of Coulomb gauge, which leads to singular propagators
for $A_0$ and the ghosts. In appendix D, we explain a more
well-defined way of performing Coulomb-gauge calculations which
avoids any $\delta(0)$ terms in the calculation but nevertheless
gives the same results, justifying our naive cancellations.}

The propagator for $A^\alpha$ that follows from \quadratic\ is
\eqn\propbiga{
\langle
A^\alpha_{ab}(t) A^\beta_{cd}(t') \rangle = \delta^{\alpha
\bar{\beta}} \Delta^{ad,cb}_{j_\alpha}(t-t',\alpha), }
where $\Delta$ is a periodic function of time given for $0 \le t \le
\beta$ by
\eqn\defdelta{ \Delta_{j}(t,\alpha) 
\equiv {e^{i \alpha t} \over 2(j+1)}
\left( {e^{-(j+1)t} \over 1 - e^{i \alpha \beta} e^{-(j+1) \beta}}
- {e^{(j+1)t} \over 1 - e^{i \alpha \beta} e^{(j+1) \beta}}
\right) \; . }
Here, $\alpha$ is shorthand for $\alpha \otimes 1 - 1
\otimes \alpha$, and a term $\alpha^n \otimes \alpha^m$ in the
expansion of $\Delta$ should be understood to carry indices
$(\alpha^n)^{ad}(\alpha^m)^{cb}$ in \propbiga.

\subsec{Regularization and counterterms}

The two-loop diagrams that we are required to evaluate contain
divergences, as we should expect from gauge theory in four
dimensions. To deal with these, we could apply dimensional
regularization, evaluating the partition function on $S^3 \times S^1
\times \IR^d$, and using a modified minimal subtraction procedure to
obtain finite and gauge-independent results. In practice, to
overcome technical difficulties with this regularization scheme, we
will use the  regularization scheme of \firstord\ involving momentum
cutoffs, which requires adding to the action counterterms designed
to yield results that are completely equivalent to those of dimensional
regularization.\foot{More precisely, the results are equivalent to 
dimensional regularization in a gauge where $3+d$ components of the
gauge field participate in the Coulomb gauge condition \gf.  This is 
related to the more general approach of split-dimensional 
regularization \refs{\splitone,\splittwo,\splitthree,\splitfour}, 
in which the number of degrees of freedom that do not participate 
in the Coulomb gauge condition is also varied.} This is
reviewed in detail in section 3. For our calculation, the
counterterm vertices that contribute are\foot{As indicated, it is
only the $SU(N)$ part of the gauge field $A_i^{SU(N)} \equiv A_i - {1
\over N} \tr(A_i)$ that can be present in the counterterms since
the $U(1)$ part of the theory is free.}
\eqn\cterm{\eqalign{ S_{ct}
&= \lambda \int_0^\beta dt \int_{S^3} \tr\left(A^{SU(N)}_i (Z_0  - Z_1
\partial_j^2 - Z_2 D_\tau^2) A^{SU(N)}_i\right) \cr &= \lambda
\int_0^\beta dt \; \tr\left(A^{\bar \alpha}(Z_0 + Z_1 (j_\alpha+1)^2 - Z_2
D_\tau^2) A^\alpha\right) \cr & - {\lambda \over N} \int_0^\beta dt \;
\tr\left(A^{\bar \alpha})(Z_0 + Z_1 (j_\alpha+1)^2 - Z_2 D_\tau^2\right)
\tr\left(A^\alpha\right), }}
where $\lambda \equiv g_{YM}^2 N$ and $Z_i$ are regulator dependent
constants that we will determine in section 3.

\subsec{One loop result}

The evaluation of the partition function at one-loop order has been
carried out in \refs{\first}.  The result is
\eqn\oneloop{ Z_{\rm{1-loop}} = \int [dU] e^{-S^{\rm eff}_{\rm
1-loop}(U)}, }
where
\eqn\aaa{ S^{\rm eff}_{\rm 1-loop}(U) = -\sum_{n=1}^\infty {1 \over
n} z(x^n) \tr(U^n) \tr(U^{\dagger n}), }
and we define
\eqn\zdef{ z(x) \equiv {6 x^2 - 2 x^3 \over (1-x)^3} }
which is the single mode partition function for a free vector field
on $S^3$. The unitary matrix integral can be evaluated explicitly at
large $N$ to give \first \eqn\explicit{ Z_{\rm 1-loop} = e^{-\beta
F_{\rm 1-loop}} = \prod_{n=1}^\infty {1 \over 1 - z(x^n)}. } The
function $z$ increases monotonically from 0 to $\infty$ as $x$
increases from 0 to 1 (i.e. as the temperature increases from 0 to
$\infty$), so the $n=1$ term in the product leads to a divergence as
the temperature is increased to the critical value $x_{c,0}$ such
that $z(x_{c,0})=1$, or
\eqn\hagtempfree{T_{c,0} R_{S^3} = (\ln(2+\sqrt{3}))^{-1} \approx 0.75933.}
This is the Hagedorn temperature of the large $N$ free theory (and
of the interacting theory in the small volume limit). For higher
temperatures (at finite $N$) there is a different saddle point
dominating the path integral, and $Z$ behaves as $\exp{(-\beta N^2
f(T))}$.

\subsec{Two loop calculation}

At two-loop order, the partition function is given by
\eqn\twoloop{
Z_{\rm 2-loops} = \int [dU] e^{-S^{\rm eff}_{\rm 1-loop}(U)-S^{\rm
eff}_{\rm 2-loop}(U)}, } where \eqn\sefftwo{\eqalign{ e^{-S^{\rm
eff}_{\rm 2-loop}} &= \langle e^{-S_{\rm int}} \rangle_{\rm 2-loop}
\cr & = \exp\left( -\langle S_4 \rangle +\half  \langle S_3 S_3
\rangle - \langle S_{ct} \rangle \right). }} Here, the expectation
values are evaluated in the free theory with fixed $\alpha$.

The correlators in \sefftwo\ contribute to the partition function in
two different ways \svv. First, as shown in \first, the planar
diagrams give a contribution of the form
\eqn\planars{ S_{pl}^{\rm eff} = \beta g_{YM}^2 N \sum_{n=1}^{\infty}
f_n(x)
\tr(U^n) \tr(U^\dagger {}^n) + \beta g_{YM}^2 \sum_{n \ge m > 0}
f_{nm}(x) \{ \tr(U^n) \tr(U^m) \tr(U^{-n-m}) + c.c.\} \; . }
The three-trace terms do not contribute to the partition function at
order $\lambda$, so they can be ignored. On the other hand, the
double-trace terms modify the Gaussian integral \oneloop\ (the path
integral is Gaussian in the variables $u_n = \tr(U^n)/N$), and
result in order $\lambda$ corrections to the denominators in
\explicit.

Since \explicit\ is actually the sub-leading contribution to the
large $N$ free energy (the leading ${\cal O}(N^2)$ contribution,
coming from the action evaluated on the saddle-point, vanishes
here), there are also contributions at the same order arising from
{\it non-planar} two-loop diagrams. These are independent of $U$
(they have a single index loop) and give a temperature-dependent
prefactor to the infinite product in \explicit,
\eqn\nps{ S_{np}^{\rm eff} = \beta g_{YM}^2 N F^{np}_2(x) \; . }

Thus, in terms of the functions $f_n(x)$ and $F_2^{np}(x)$ defined
in \planars\ and \nps, the final result for the two-loop partition
function is
\eqn\twoform{\eqalign{ Z &= e^{- \lambda \beta F^{np}_2(x)} \int
[dU] e^{\sum_n {1 \over n} (z(x^n) - \lambda \beta n f_n(x)) \tr(U^n)
\tr(U^\dagger {}^n) + \cdots} \cr &=  e^{- \lambda \beta F^{np}_2(x)}
\prod_{n=1}^\infty {1 \over 1 - z(x^n) + \lambda n \beta f_n(x)} +
{\cal O} (\lambda^2). }}
Expressed in terms of the
correction to the free-energy, we have
\eqn\fcorr{ \delta F = \lambda \left(F_2^{np}(x) + \sum_{n=1}^\infty
{n f_n(x) \over 1-z(x^n)} \right). }
From \twoform, we see that the corrected
partition function will diverge when
\eqn\xdiv{1 - z(x) + \lambda \beta f_1(x) = 0,}
so we find that the critical value of $x$ shifts by
\eqn\xshift{ \delta x_c = \lambda \beta_{c,0} {f_1(x_{c,0}) \over z'(x_{c,0})} }
or, equivalently, the critical temperature shifts by
\eqn\tshift{ \delta T_c = \lambda T_{c,0} {f_1(x_{c,0}) \over x_{c,0} z'(x_{c,0})}
\; . }
It remains to evaluate $f_n(x)$ and $F^{np}_2(x)$ by evaluating the
planar and non-planar two-loop diagrams plus one-loop counterterm
diagrams. We do this in section 4, but first we must discuss our
regularization procedure and determine the necessary counterterms.

\newsec{Regularization and counterterms}

The regularization procedure that we use was described in detail in
\firstord, so we only summarize it briefly here. The central idea is
to cut off angular momentum sums in such a way that our calculations
are as simple as possible. Such a scheme will in general break gauge
and Lorentz invariance but, by choosing the right set of local
counterterms in the action at the cutoff scale $M$,
we can ensure that
both are restored in the theory far below the cutoff. Specifically,
we will choose counterterms so that the results for any correlator
match precisely, in the limit where the cutoff is removed, with
those obtained using dimensional regularization.

Though it would be conceptually simpler to perform our calculations
directly in dimensional regularization, this presents technical
challenges for the required diagrams that we have not been able to
overcome.  On the other hand, determining the appropriate set of
counterterms to ensure the equivalence of a more general cutoff
scheme with dimensional regularization requires, at least to low
orders in perturbation theory, the evaluation of only a few simple
diagrams using both methods and a comparison of the results.

The regulator that we employ requires the insertion of a damping
factor $R(\sqrt{p^2/M^2})$ for each $A_i$ propagator, where
$p^2=p_ip_i$ ($i=1,2,3$) is the magnitude of the spatial momentum
and $R$ is a function satisfying $R(0)=1$, $R'(0)=0$, and
$R(x\rightarrow\infty)=0$ \foot{Note that we do not use a regulator
for $A_0$ or ghost lines, which we have already integrated out
explicitly. As described in appendix D, the cancelling $\delta(0)$
terms that appear in integrating out these fields are not physical
UV divergences and may be regulated by choosing a more general
gauge.}. The regulator preserves rotational invariance, but breaks
Lorentz and gauge-invariance, so we should include all possible
counterterms which are local, rotationally invariant, and have
dimension four or less. Since all counterterms will be at least of
order $\lambda$, only quadratic counterterms (giving rise to
one-loop diagrams) can contribute to the partition function at order
$\lambda$. Moreover, only the $A_i$ propagators are
temperature-dependent with our choice of gauge. Thus, only the
counterterms appearing in \cterm\ can contribute to our
result\foot{Note that a possible term of the form $\del_i A_i \del_j
A_j$ will not contribute due to the Coulomb gauge choice.}.

In order to determine the constants $Z_0$, $Z_1$, and $Z_2$, we need
to evaluate at least three simple one-loop correlators to which
$Z_0$, $Z_1$ and $Z_2$ contribute in different linear combinations,
and choose the $Z$'s so that the results match with the same
correlators evaluated in dimensional regularization, with a minimal
subtraction scheme to be described below.

It is important to note that for our calculation on the sphere, the
$Z_0$ counterterm actually combines three separate local covariant
structures,
\eqn\counters{\int d^4 x \sqrt{g} \left( Z_0^{flat} \tr(A_i A_i) + Z_0'
{\cal R} \tr(A_i A_i) + Z_0'' {\cal R}_{ij} \tr(A_i A_j)\right),}
the last two involving the Ricci scalar and Ricci tensor built from
the metric. If $Z_0'$ or $Z_0''$ are nonzero, the $\tr(A_i A_i)$
counterterm on the sphere will be different than the one in flat
space, with coefficients proportional to $1/R_{S^3}^2$ in addition to the
flat space coefficient $Z_0^{flat}$
proportional to $M^2$. Thus, it is important
that the calculation used to determine $Z_0$ be performed on the
sphere, as we will do in section 3.2.

On the other hand, the counterterms involving $Z_1$ and $Z_2$ are
already dimension 4 operators, so there is no allowed structure
involving the spatial curvature that reduces to these. Consequently,
it is enough to study flat-space correlators to determine $Z_1$ and
$Z_2$, and we turn to this presently.

\subsec{Curvature-independent counterterms}

The $Z_1$ and $Z_2$ counterterms were already determined in
\firstord, but for completeness, we review the essential parts of
that calculation here. As we suggested above, it is simplest to
determine the curvature-independent counterterms by a flat space
calculation. To determine $Z_1$, we will compute
\eqn\firstcor{ \langle A_i(0,p) A_j(0,-p) \rangle_{p^2
\delta_{ij}}^{1PI} }
i.e. the term proportional to $p^2 \delta_{ij}$
in the 1PI two-point function of $A_i$, while to determine $Z_2$, we
will compute
\eqn\secondcor{ \langle A_i(\omega, 0) A_j(-\omega,0)
\rangle_{\omega^2}^{1PI} \; , }
with $\omega$ that component of the momentum which does not
participate in the Coulomb gauge constraint. In each case, we
compute the result (to order $\lambda$) using our regulator and the
counterterm with undetermined coefficient, then repeat the
calculation in dimensional regularization with a modified minimal
subtraction scheme.  We finally determine the counterterm by demanding that
the two calculations agree.

\vskip 0.1 in
\noindent
{\bf Dimensional regularization}
\vskip 0.1 in

We will begin with the calculation in dimensional regularization. We
generalize our Coulomb gauge by assuming that $3+d$ components of the
gauge field participate in the Coulomb gauge condition. With this
choice, we may write the quadratic action as
\eqn\quadr{S_2 = \int d^{d+4} x \; \tr \left\{{1 \over 2} \dot{A}_i
\dot{A}_i + {1 \over 2} \partial_j A_i \partial_j A_i + {1 \over 2}
\partial_i A_0 \partial_i A_0 + \partial_i \bar{c} \partial_i c
\right\}.}
The interaction terms for Euclidean Yang-Mills theory on $\IR^{d+4}$
in the Coulomb gauge include a cubic action
\eqn\cubac{ S_3 = g_{YM} \int
d^{d+4} x \; \tr \left\{ -i \partial_i A_j [A_i, A_j] -i \dot{A}_i
[A_0, A_i] + i \partial_i A_0 [A_0, A_i]  - i \partial_i \bar{c}
[A_i ,c] \right\} }
and quartic terms
\eqn\quart{S_4 = g_{YM}^2
\int d^{d+4} x \; \tr \left\{ -{ 1 \over 4} [A_i ,
A_j]^2 - {1 \over 2} [A_0 , A_i]^2 \right\} \; .}
From the quadratic action, we may derive propagators (suppressing
the color indices)
\eqn\flatprops{\eqalign{ \langle A_i(\nu, k) A_j(-\nu, -k) \rangle &
\equiv \Delta_{ij}(\nu, k)  = {k^2 \delta_{ij} - k_i k_j \over
k^2(\nu^2 + k^2)}, \cr \langle A_0(\nu, k) A_0(-\nu, -k) \rangle &=
{1 \over k^2}, \cr \langle c(\nu, k) \bar{c}(-\nu, -k) \rangle &= {1
\over k^2}. }}

\fig{Diagrams that contribute to the two-point function $\langle
A_iA_j\rangle$.  Dashed (arrowed) lines denote $A_0$ (ghost)
propagators, and solid lines denote $A_i$ propagators.}
{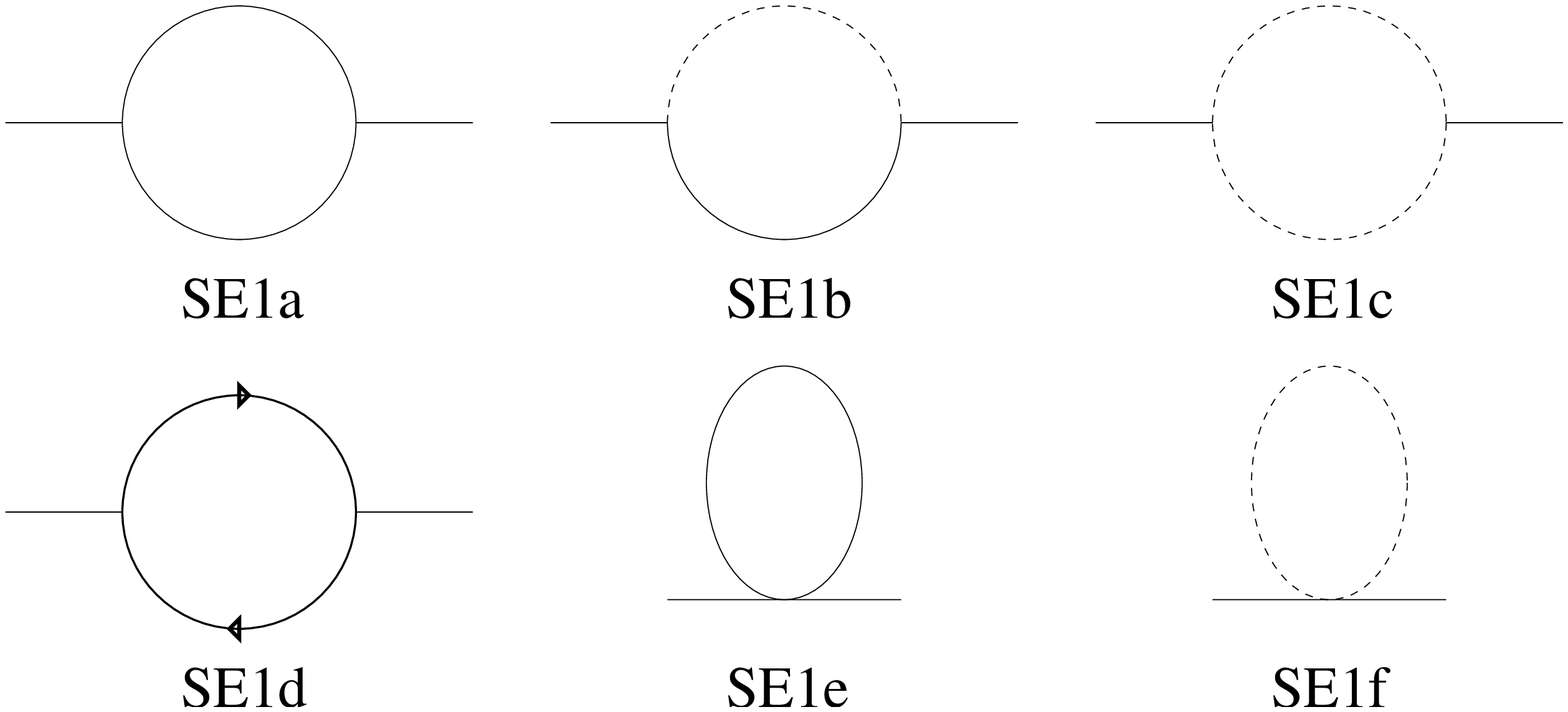}{5.0truein}\figlabel{\twopointfig}

The quantities
\firstcor\ and \secondcor\ are both obtained from the 1PI two-point
function of the gauge field. At one-loop, this gets contributions from
the six diagrams of figure \twopointfig.  Note, however, that diagrams
SE1e and SE1f, which involve only quartic vertices, do not depend on
the external momenta and hence do not contribute to \firstcor\ or
\secondcor.  In addition, the $A_0$ loop diagram SE1c cancels with the
ghost loop diagram SE1d{\foot{Actually, these two diagrams only cancel
up to a $\delta(0)$ divergence of the usual sort associated to $A_0$
lines.  As discussed in Appendix D, this divergence is completely
unphysical and is not important for any of our results.}} so we need
only focus our attention on the first two, SE1a and SE1b.  For these,
we find

\eqn\aaa{\eqalign{ &-{1 \over 2} \langle A_i(\omega,p) A_j(-\omega,
-p) \rangle_{SE1a}^{1PI} \cr & \qquad =  \int {d \nu d^{d+3} k \over
(2 \pi)^{d+4} }  \left\{ - \Delta_{ij} [(p+k) \cdot \hat{\Delta}
\cdot (p+k)] + [(p+k)\cdot \hat{\Delta} \cdot \Delta]_i (2k-p)_j
\right. \cr & \left. \qquad \qquad + [(p+k)\cdot \hat{\Delta} \cdot
\Delta]_j (2k-p)_i  + [(2p-k) \cdot \Delta]_i [(p+k) \cdot
\hat{\Delta}]_j + \Delta_{kl} \hat{\Delta}_{kl} k_i (p-2k)_j
\right\}, }}
where $\Delta \equiv \Delta(\nu,k)$ and $\hat{\Delta} \equiv
\Delta(\omega-\nu, p-k)$, and
%
\eqn\aa{\eqalign{ -{1 \over 2} \langle A_i(\omega,p) A_j(-\omega,
-p) \rangle_{SE1b}^{1PI} &=  \int {d \nu d^{d+3} k \over (2
\pi)^{d+4} }  \left\{ - \Delta_{ij} {(\omega+ \nu)^2 \over (p-k)^2}
\right\}. }}
It is straightforward to insert explicit expressions
for the propagators and expand to order $p^2$ or $\omega^2$. In the
latter case, we find\foot{We have used the fact that $k_i k_j$
inside the integral may be replaced by $k^2 \delta_{ij}/(3+d)$ if
the external momentum $p$ is set to zero.}
\eqn\scor{\eqalign{ -{1 \over 2}  \langle A_i(\omega, 0)
A_j(-\omega,0) \rangle_{\omega^2}^{1PI} &= \omega^2 \delta_{ij} {d+2
\over d+3} \left\{-I_{0,1}(0) + 2  I_{0,3}(0) - 8  I_{1,4}(0)
\right\}, }}
where we define
\eqn\idef{\eqalign{ I_{m,n}(a) &\equiv \int {d^{3+d} k \; d \nu
\over (2 \pi)^{4+d}} {\nu^{2m} k^{2n-2m-2} \over (k^2 + a^2)
(\nu^2 + k^2)^n} \cr &={a^d \over 2^{4+d}\mu^d(\pi)^{5+d \over
2}} { \Gamma(n-m - {1 \over 2} ) \Gamma(m+ {1 \over 2}) \Gamma(1+ {d
\over 2}) \Gamma(-{d \over 2}) \over \Gamma(n) \Gamma({3 \over 2} +
{d \over 2})} \cr & =
\frac{1}{4\pi^3}\frac{\Gamma\left(n-m-\frac{1}{2}\right)
\Gamma\left(m+\frac{1}{2}\right)}{\Gamma(n)}\left(\ln\left(\frac{\mu}{a}\right)
+\left[\frac{1}{\epsilon}+\frac{1}{2}\ln(\pi)-\frac{\gamma}{2}+1\right]
\right)\;.
}}
Here, $\mu$ is the regularization scale introduced, as usual, to
keep $I_{m,n}(a)$ dimensionless as $d$ is varied.  In the last line,
we have set $d = - \epsilon$ and expanded for small $\epsilon$. The
integrals for $a=0$ in \scor\ contain infrared divergences in
addition to the UV divergences, so rather than comparing these
directly between the two schemes, we write $I(0) = I(a) +
\{I(0)-I(a)\}$, and, noting that the expression in curly brackets
contains no UV divergence and must agree between the two schemes,
focus on the $I(a)$ term and compare this between the two schemes to
determine the counterterms.

For our calculations, it is convenient to choose a minimal subtraction
scheme which sets the combination in square brackets in \idef\ to
zero. With this choice, the only non-vanishing contributions to
\scor\ are terms proportional to $\ln(\mu/a)$ and
terms for which an $\epsilon$ in the expansion of the
$d$-dependent coefficients multiplies the $1 \over \epsilon$ part of
some $I_{m,n}$. The result for \secondcor\ is thus
\eqn\scorfin{\eqalign{ -{1 \over 2} \langle A_i(\omega, 0)
A_j(-\omega,0) \rangle_{\omega^2}^{1PI} &=
\omega^2\delta_{ij}\left(\frac{1}{48\pi^2}-\frac{1}{8\pi^2}\ln\left(\frac{\mu}{a}\right)+\{
{\rm UV \; finite }\}\right),}}
%
%
where the UV finite term denotes the difference between the original
integral and the IR regulated version{\foot{Note that this term will
contain a dependence on the IR cutoff $\ln(a)$, which exactly
cancels that explicitly written.}}. Following the same steps, we find
that
\eqn\fcorfin{\eqalign{ -{1 \over 2} \langle A_i(0,p) A_j(0,-p)
\rangle_{p^2 \delta_{ij}}^{1PI} &= p^2 \delta_{ij} \left\{-{4d^2+22d+20
\over (d+3)(d+5)} I_{0,2}(0) + {d^2 + 3d - 6 \over (d+3)(d+5)}
I_{1,1}(0)  \right. \cr & \left.  \qquad  \qquad - {8(2+d) \over
(d+3)(d+5)} I_{0,4}(0) + 
{2d^2+14d+4 \over (d+3)(d+5)}
I_{0,3}(0)\right\} \cr & =
p^2\delta_{ij}\left(\frac{29}{240\pi^2}-\frac{1}{8\pi^2}\ln\left(\frac{\mu}{a}\right)+\{
{\rm UV \; finite} \}\right). }}

\vskip 0.1 in
\noindent
{\bf Cutoffs and Counterterms}
\vskip 0.1 in

We now reevaluate the correlators using the regulated momentum
integrals together with counterterms, choosing the counterterm
coefficients so that the results match with those of \scorfin\ and
\fcorfin. Our regularization scheme employs a cutoff only for the
$A_i$ lines, so the regulated expressions for diagrams SE1a and SE1b
follow by setting $d=0$ and replacing $\Delta_{ij}(\nu, k) \to
\Delta_{ij}(\nu,k) R(k/M)$ in \aaa\ and \aa. Our results may be expressed in terms of
the following basic integrals:
\eqn\regdefs{\eqalign{ \ln\left(\frac{{\cal{A}}_nM}{\mu}\right)&
\equiv \int_0^{\infty}\,dq\,
\frac{\sqrt{q^2}R^n(q)}{q^2+\frac{\mu^2}{M^2}}, \cr
C_n&\equiv \frac{1}{4\pi^2}\int_0^{\infty}\,dq\,q R^n(q),\cr
F_2&\equiv \frac{1}{4\pi^2}\int_0^{\infty}\,dq\,q\,R(q)R^{\prime\prime}(q).\cr
}}
Again, we start with \secondcor, for which we can set $p=0$. In this
case, the contributions from \aaa\ and \aa\ now include regulator
factors of $R^2(k/M)$ and $R(k/M)$, respectively. Starting from
\scor\ with $d=0$, we can evaluate the integrals over $\nu$ to get
\eqn\imn{I_{m,n}(a) \to {1 \over 2 \pi} {\Gamma(n-m-{1 \over 2})
\Gamma(m+{1 \over 2}) \over \Gamma(n)} \int {d^3 k \over (2 \pi)^3}
{1 \over k (k^2 + a^2)}}
and then insert the regulators. Including also the counterterm
contribution, we find
\eqn\scorcc{\eqalign{ -{1 \over 2} \langle A_i(\omega, 0)
A_j(-\omega,0) \rangle_{\omega^2}^{1PI} &= \omega^2 \delta_{ij}
\left\{ Z_2 + \int {d^3 k \over (2 \pi)^3} {1 \over k (k^2 + a^2)}
\left(-{1 \over 3} R(k/M)+ {1 \over 12} R^2(k/M)\right) \right\} \cr
&=  \omega^2 \delta_{ij}\left\{ Z_2 -{1 \over 6 \pi^2} \ln
\left({{\cal A}_1 M \over a} \right) + {1 \over 24 \pi^2} \ln
\left({{\cal A}_2 M \over a} \right) \right\}. }}
For the other correlator \firstcor, we again need a regulator factor
$R(k/M)$ for the contribution \aa\ from SE1b, and this time the factor
$R(k/M)R(|p-k|/M)$ for the contribution \aaa\ from SE1a. It is
important to take the expansion of $R(|p-k|/M)$ in powers of $p$
into account in order
to correctly obtain the terms proportional to $p^2 \delta_{ij}$ in
\aaa.  The effect of this is trivial for logarithmic divergences as
only the first term, $R(k/M)$, contributes there. However, the expression
\aaa\ has a quadratic divergence, for which subleading terms
in the expansion become important.  Working everything out carefully,
we find
\eqn\fcorcc{\eqalign{ -{1 \over 2} \langle A_i(0,p) & A_j(0,-p)
\rangle_{p^2 \delta_{ij}}^{1PI} = 
p^2\delta_{ij}\left\{Z_1+\int\,\frac{d^3k}{(2\pi)^3}\frac{1}{k(k^2+a^2)}
\left(\frac{1}{5}R(k/M)-\frac{2}{5}R(k/M)^2\right)\right\}\cr
&\qquad -4\left[\int\frac{d^3k\,d\nu}{(2\pi)^4}
\frac{k^2k_ik_j}{(p-k)^2(k^2+\nu^2)([p-k]^2+\nu^2)}R(k/M)R(|p-k|/M)
\right]_{p^2\delta_{ij}}\cr
&= p^2
\delta_{ij}\left\{ Z_1 +{1 \over 10 \pi^2} \ln \left({{\cal A}_1 M
\over a} \right) - {9 \over 40 \pi^2} \ln \left({{\cal A}_2 M \over
a} \right) -{F_2 \over 15} -\frac{1}{40\pi^2} \right\}, }}
where the first line contains the contribution from \aa\ as well as
the purely logarithmically divergent terms 
in \aaa\ and the second line contains
the only term in \aaa\ for which $R(|p-k|/M)$ must be expanded in $p$ 
beyond the leading order.
Finally, demanding that \fcorcc\ and \scorcc\ agree with \fcorfin\
and \scorfin, we must have
\eqn\cts{\eqalign{ Z_1 &= {1 \over 8\pi^2}
\ln\left({M\over \mu}\right) + {1\over 15} F_2 + {7 \over 48 \pi^2}
- {1\over 40\pi^2} \ln\left({{\cal A}_1^4 \over {\cal
A}_2^9}\right), \cr Z_2 &= {1\over 8\pi^2} \ln\left({M\over
\mu}\right) + {1\over 48\pi^2} - {1\over 24\pi^2} \ln\left({{\cal
A}_2 \over {\cal A}_1^4}\right). \cr }}

\subsec{Curvature dependent counterterm}

As we discussed above, the $\tr(A_i A_i)$ counterterm has a
coefficient which depends on the spatial curvature, so we need to
determine $Z_0$ by a direct calculation on $S^3$, though we can work
on $S^3 \times \IR^1$ rather than $S^3 \times S^1$. Thus, to find
$Z_0$ we compute the two-point function of a mode of $A_i$ on the
sphere, again demanding that a calculation using regulator functions
and counterterms matches with the result using dimensional
regularization (this time gauge theory on $\IR^{1+d} \times S^3$).
Specifically, we will compute
\eqn\sphtwo{ R=\sum_{m,m',\epsilon} \langle
\bar{A}^{j=1,m,m',\epsilon}(\omega=0)
A^{j=1,m,m',\epsilon}(\omega=0) \rangle_{1PI} \; , }
the 1PI two-point function of the lowest total angular momentum mode
of $A_i$, summed over polarizations $\epsilon$ and angular momentum
states $m,m'$. We begin with the calculation in dimensional
regularization.

\vskip 0.1in
\noindent
{\bf Dimensional regularization}
\vskip 0.1 in

To define the dimensionally regularized theory, we choose
\eqn\dimregact{S_{Euc} = \int d^{d+1} x d^3 y \tr \left({1 \over 4}
F_{IJ} F_{IJ} \right) \; ,}
where $y$ are the coordinates on $S^3$ and $x$ are the coordinates
on $\IR^{d+1}$ with $d=-\epsilon$, and for simplicity we take the
regularization scale $\mu$ to be unity (in units of $R_{S^3}^{-1}$),
\eqn\dimregsc{\mu=1.}
We denote by $A_a$ the components of the gauge field in the $d$
directions, while $A_0$ and $A_i$ denote as before the components of
the gauge field in the time and sphere directions. We work in the
Coulomb gauge, now extended to involve $d+3$ components\foot{For
this part of the calculation, it is actually much simpler to choose
a gauge for which only the sphere components of the gauge field
participate in the Coulomb gauge condition, but we must make the
present choice to be consistent with the conventions of \firstord\
and of the previous subsection, used to compute $Z_1$ and $Z_2$.}
\eqn\dimreggauge{
\partial_i A_i + \partial_a A_a = 0 \; .}
Since the divergence of $A_i$ is no longer set to zero, the expansion
of modes on the sphere now becomes
\eqn\newexp{\eqalign{ A_0 &= \sum_{\alpha} a^\alpha S^\alpha, \cr
A_a &= \sum_{\alpha} {\cal A}^\alpha_a S^\alpha, \cr
A_i &= \sum_{\beta} A^\beta V_i^\beta + \sum_{\alpha} {1
\over j_\alpha (j_\alpha + 2)} \partial_a {\cal A}^\alpha_a
\partial_i S^\alpha \; . }}
Note that the Coulomb gauge condition determines the coefficients of
$\nabla S$ in the expansion of $A_i$ in terms of the ${\cal{A}}_a$ modes.
With these expansions, we find that the quadratic action may be
written as
\eqn\squad{\eqalign{ S_2 &= \int d t d^d x \tr \left( {1
\over 2} A^\alpha (-\partial_t^2 - \partial_a^2 + (j_\alpha + 1)^2)
A^\alpha \right. \cr & \qquad + {1 \over 2} a^\alpha ( -
\partial_a^2 + j_\alpha (j_\alpha + 2)) a^\alpha \cr & \left. \qquad
+ {1 \over 2} {\cal A}_a^\alpha (-\partial_t^2 -
\partial_c^2 + j_\alpha(j_\alpha + 2))(\delta^{ab} - {\partial_a
\partial_b \over j_\alpha (j_\alpha + 2)}) {\cal A}^\alpha_b \right)
\; .}}
From these, we determine the propagators to be
\eqn\props{\eqalign{ \langle a^\alpha(\omega,k)
a^\beta(\tilde{\omega},\tilde{k}) \rangle  &=  (2 \pi)^{d+1}
\delta(\omega + \tilde{\omega}) \delta^d(k + \tilde{k})
{\delta^{\alpha \bar{\beta}} \over k^2 + j_\alpha(j_\alpha + 2)},
\cr \langle A^\alpha(\omega,k) A^\beta(\tilde{\omega},\tilde{k})
\rangle &=  (2 \pi)^{d+1} \delta(\omega + \tilde{\omega}) \delta^d(k
+ \tilde{k}) {\delta^{\alpha \bar{\beta}} \over \omega^2 + k^2 +
(j_\alpha + 1)^2}, \cr \langle {\cal A}_a^\alpha(\omega,k) {\cal
A}_b^\beta(\tilde{\omega},\tilde{k}) \rangle  &=  (2 \pi)^{d+1}
\delta(\omega + \tilde{\omega}) \delta^d(k + \tilde{k})
{\delta^{\alpha \bar{\beta}} \over \omega^2 + k^2 +
j_\alpha(j_\alpha + 2)} (\delta_{ab} - {k_a k_b \over k^2 + j_\alpha
(j_\alpha + 2)}). }}
There are now additional interaction terms involving ${\cal A}$. For
our calculation of the two-point function of $A_i$, we will need the
cubic terms with at least one power of this field, and the quartic
terms with either two or four powers of ${\cal A}$. In addition to
those terms which appear for $d=0$, listed in (A.2) and (A.3), the
new terms of this form are:
\eqn\cubicn{\eqalign{ {\cal L}_{Aa{\cal A}} = & g_{YM} \tr \left( -i
\partial_0 \partial_a {\cal A}_a^\alpha [a^\beta, A^\gamma]
{C_{\beta \gamma \alpha} \over j_\alpha (j_\alpha + 2)} - i
\partial_0 A^\alpha [a^\beta, \partial_a {\cal A}_a^\gamma]
{C_{\beta \alpha \gamma} \over j_\alpha (j_\alpha + 2)} \right), \cr
{\cal L}_{A A{\cal A}} = & g_{YM} \tr \left( -i[A^\alpha, \partial_a
A^\beta] {\cal A}_a^\gamma D^{\alpha \beta \gamma} +
i{(j_\alpha+1)^2 - (j_\beta +1)^2 \over j_\gamma (j_\gamma + 2)}
A_\alpha A_\beta \partial_a {\cal A}_a^\gamma D^{\alpha \beta
\gamma} \right) \cr 
{\cal L}_{A {\cal A} {\cal A}} = & g_{YM} \tr \left( 2i
{\cal A}_a^\alpha A^\gamma {\cal A}_a^\beta C^{\alpha \gamma \beta}
- i [\partial_b {\cal A}_b^\alpha , \partial_a A^\beta] {\cal
A}_a^\gamma {C^{\gamma \beta \alpha} \over j_\alpha (j_\alpha + 2)}
\right. , \cr &\left. - i[A^\alpha, \partial_a \partial_b {\cal
A}_b^\beta]{\cal A}_a^\gamma {C^{\gamma \alpha \beta} \over j_\beta
(j_\beta + 2)}  - i A^\alpha \partial_a {\cal A}_a^\beta \partial_b
{\cal A}_b^\gamma {(j_\alpha + 1)^2 \over j_\beta (j_\beta + 2)
j_\gamma (j_\gamma + 2)} C^{\beta \alpha \gamma} \right), }}
and
\eqn\quarticn{\eqalign{ {\cal L}'_4 =& g_{YM}^2 \tr \left( -{1 \over 2} [{\cal
A}_a^\alpha, A^\beta][{\cal A}_a^\gamma, A^\delta] D^{\beta \delta
\lambda} B^{\alpha \gamma \bar{\lambda}} -{1 \over 2} [A^\alpha,
\partial_a {\cal A}_a^\beta][A^\gamma, \partial_b {\cal A}_b^\delta]
D^{\alpha \gamma \lambda} \hat{B}^{\beta \delta \bar{\lambda}}
\right. \cr & \left. -{1 \over 2} [A^\alpha, \partial_a {\cal
A}_a^\beta][\partial_b {\cal A}_b^\gamma, A^\delta] C^{\lambda
\alpha \gamma} C^{\bar{\lambda} \delta \beta} -{1 \over 2}
[A^\alpha,A^\beta][\partial_a {\cal A}_a^\gamma, \partial_b {\cal
A}_b^\delta] {C^{\lambda \alpha \gamma} C^{\bar{\lambda} \beta
\delta} \over j_\gamma(j_\gamma + 2) j_\delta (j_\delta + 2)}
\right), }}
where $B$, $C$ and $\hat{B}$ are defined in appendix B.

\fig{New diagrams with ${\cal A}_a$ propagators that contribute to the
two-point function $\langle A_iA_j\rangle$.  Dotted (dashed)
lines denote ${\cal{A}}_a$ ($a$) propagators.}
{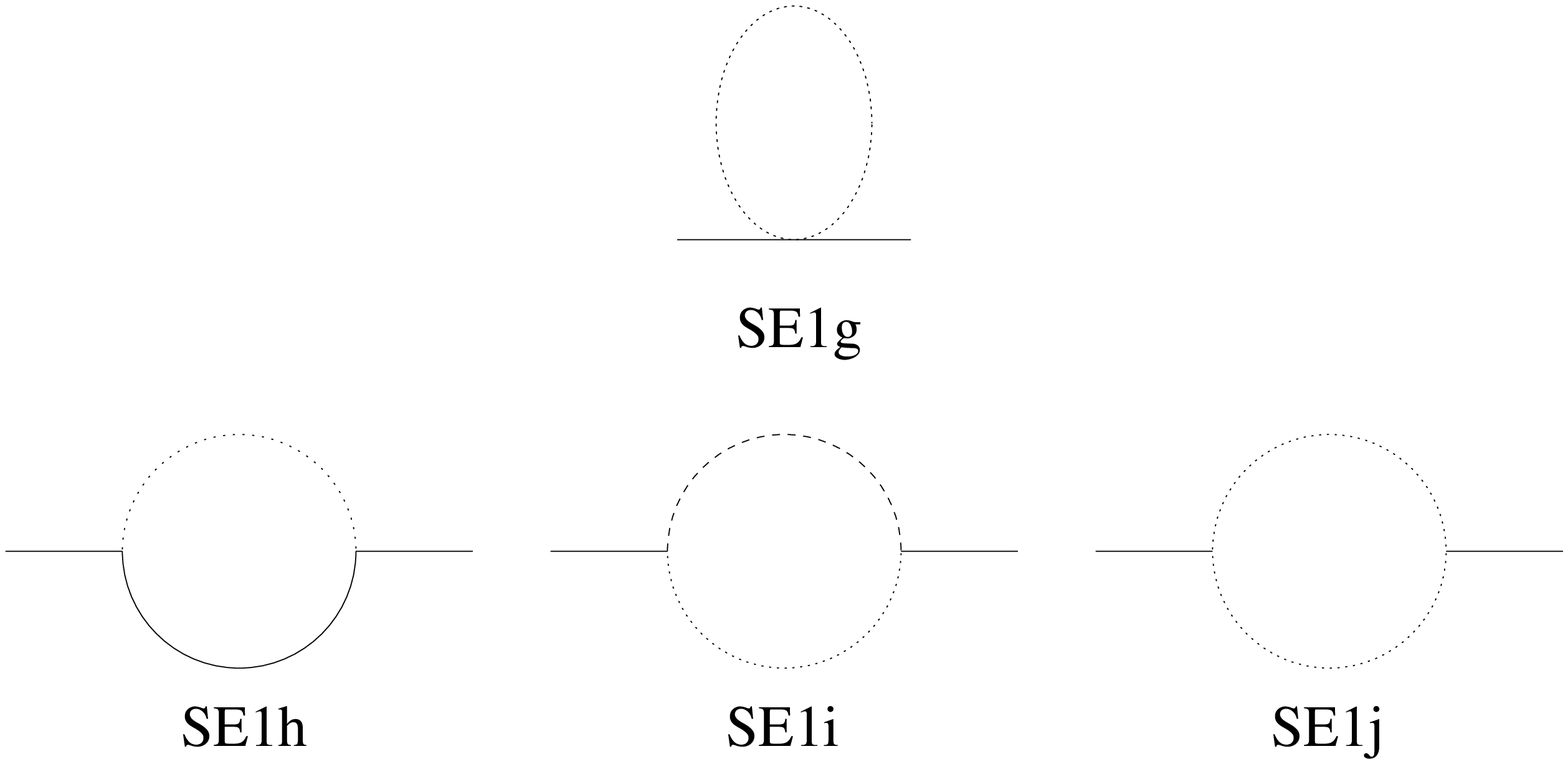}{5.0truein} \figlabel{\newtwopointfig}

We are now ready to calculate \sphtwo, setting the external momentum
in the $d$ directions to zero also.  In addition to the diagrams of
figure \twopointfig, the new interaction terms \cubicn\ and \quarticn\
give rise to the diagrams of figure \newtwopointfig.  Note that
because the latter contain internal ${\cal{A}}$ lines, they yield
contributions to the two point function only through the
multiplication of $1/\epsilon$ poles (from sums which are
logarithmically divergent at $\epsilon=0$) and prefactors proportional
to $\epsilon$.

For the $n$'th diagram, we define $\Pi^{\alpha \beta}_n$ to be the
correlator for two arbitrary modes $A^\alpha$ and $A^\beta$. 
This has a contribution proportional
to $\delta(0)$, which we denote by $\delta_n$ (these cancel 
between the diagrams, and the naive
cancellation may be justified by the method in appendix D), and
a finite contribution.
After setting $\beta = \bar{\alpha}$, choosing $j_\alpha = 1$ and summing
over the external $m$'s and $\epsilon$'s, we denote the finite
contribution by $G_n$.
Some useful formulae may be found in appendix C.

We begin with the diagrams involving quartic vertices. Diagram SE1e
gives
\eqn\forpione{ \Pi_{SE1e} = 2(D^{\alpha \gamma \lambda} D^{\beta
\bar{\gamma} \bar{\lambda}} - D^{\alpha \beta \lambda} D^{\gamma
\bar{\gamma} \bar{\lambda}}) \int {d \omega d^d k \over (2
\pi)^{d+1}} {1 \over k^2 + \omega^2 + (j_\gamma + 1)^2}.}
 From this, we find
\eqn\pone{\eqalign{ G_{SE1e} & = -{8 \over \pi^2} {1 \over  (4 \pi)^{d +
1  \over 2}} \Gamma ({1 \over 2}- {d \over 2}) \sum_{j = 1}^\infty
{j(j+2) \over (j+1)^{1-d}} \cr &=-{8 \over \pi^2} {1 \over (4
\pi)^{d +1 \over 2} } \Gamma ({1 \over 2} - {d \over 2})
(\zeta(-1-d)- \zeta(1-d)) \cr &= {4 \over \pi^2} ({1 \over \epsilon}
+ {1 \over 2} \ln(\pi) + {1 \over 2} \gamma) + {1 \over 3 \pi^2} +
{\cal O} (\epsilon). }}

We turn next to diagram SE1f
\eqn\forpitwo{\Pi_{SE1f} = -2(D^{\alpha \lambda \gamma} D^{\beta
\bar{\lambda} \bar{\gamma}} + {1 \over j_\lambda (j_\lambda + 2)}
C^{\gamma \alpha \lambda} C^{\bar{\gamma} \beta \bar{\lambda}}) \int
{d^d k d \omega \over (2 \pi)^{d+1}} {1 \over k^2 + j_\gamma
(j_\gamma + 2)}.}
This gives (at order $\epsilon^0$) a $\delta(0)$ term
\eqn\fordeltatwo{\delta_{SE1f} = -2 \delta(0) (D^{\alpha \lambda \gamma}
D^{\beta \bar{\lambda} \bar{\gamma}} + {1 \over j_\lambda (j_\lambda
+ 2)} C^{\gamma \alpha \lambda} C^{\bar{\gamma} \beta
\bar{\lambda}}) {1 \over j_\gamma (j_\gamma + 2)}}
and no remaining finite contribution,
\eqn\forrtwo{G_{SE1f} = 0 \; .}

We move now to diagram SE1g, which receives contributions from the
four quartic interaction terms in \quarticn.  The first such term
contributes
\eqn\forpithree{\Pi_{SE1g,1} = -2 D^{\alpha \beta \lambda} B^{\gamma
\bar{\gamma} \bar{\lambda}} \int {d^d k d \omega \over (2
\pi)^{d+1}} {1 \over \omega^2 + k^2 + j_\gamma (j_\gamma + 2)}(d -
{k^2 \over k^2 + j_\gamma (j_\gamma + 2)}).}
We find that the two terms in the integral actually cancel each other, so the net result is
\eqn\forrthree{G_{SE1g,1} = 0 \; .}

For the second term in \quarticn\ we find
\eqn\forpifour{\Pi_{SE1g,2} = -2 D^{\alpha \beta \lambda} \hat{B}^{\delta
\bar{\delta} \bar{\lambda}} \int {d^d k d \omega \over (2
\pi)^{d+1}} {k^2 j_\delta (j_\delta + 2) \over (\omega^2 + k^2 +
j_\delta (j_\delta + 2))( k^2 + j_\delta (j_\delta + 2))}.}
To evaluate this, we perform the integral over $\omega$ followed by
the $k$ integral. The resulting summand may be expanded in powers of
$1/j$, and since we find an overall factor of $d$, the only
contributing term is
\eqn\morepifour{-{3 d \over 2 \pi} \sum_j {1 \over (j+1)^{1-d}} \to
{3 \epsilon \over 2 \pi^2} \zeta(1+\epsilon) \to {3 \over 2 \pi^2} +
{\cal O}(\epsilon).}
Thus, this diagram gives a net contribution
\eqn\forrfour{G_{SE1g,2} = {3 \over 2 \pi^2}.}

From the third term in \quarticn, we find
\eqn\forpifive{\Pi_{SE1g,3} = 2 {C^{\lambda \alpha \gamma} C^{\bar{\lambda}
\beta \bar{\gamma}} \over j_\gamma (j_\gamma + 2)} \int {d^d k d
\omega \over (2 \pi)^{d+1}} {k^2  \over (\omega^2 + k^2 + j_\gamma
(j_\gamma + 2))( k^2 + j_\gamma (j_\gamma + 2))}.}
The evaluation of this is similar to the previous diagram, and we
find a net contribution of
\eqn\forrfive{G_{SE1g,3} = - {1 \over 2 \pi^2}.}
Finally, the last term in \quarticn\ gives no contribution.
\eqn\Rgfour{G_{SE1g,4}=0}

We now move to the cubic diagrams. Diagram SE1a gives
\eqn\forpisix{\Pi_{SE1a} = 9 \hat{E}^{\alpha \gamma \delta}
\hat{E}^{\beta \bar{\gamma} \bar{\delta}} \int {d \omega d^d k \over
(2 \pi)^{d+1}} {1 \over \omega^2 + k^2 + (j_\gamma + 1)^2}{1 \over
\omega^2 + k^2 + (j_\delta + 1)^2}.}
In this case, choosing $j_\alpha = 1 $ forces $j_\delta = j_\gamma$ by
the triangle inequality. We find
\eqn\ptwo{\eqalign{ G_{SE1a} & = {8 \over \pi^2} {1 \over  (4 \pi)^{d +1
\over 2}} \Gamma ({3 \over 2} - {d \over 2}) (\zeta(-1-d)+ 4
\zeta(1-d) - 5 \zeta(3-d)) \cr &= {8 \over \pi^2} ({1 \over
\epsilon} + {1 \over 2} \ln(\pi) + {1 \over 2} \gamma) + {47 \over 6
\pi^2} - {10 \over \pi^2} \zeta(3) + {\cal O} (\epsilon). }}

Diagram SE1b gives
\eqn\forpiseven{\Pi_{SE1b} = 2 D^{\alpha \rho \gamma} D^{\beta \bar{\rho}
\bar{\gamma}} \int {d \omega d^d k \over (2 \pi)^{d+1}} \left( {1
\over k^2 + j_\gamma (j_\gamma + 2)} -  {k^2 + (j_\rho+1)^2 \over
(k^2 + j_\gamma (j_\gamma + 2))( \omega^2 + k^2 + (j_\rho + 1)^2)}
\right).}
The first term here gives a $\delta(0)$ term
\eqn\fordeltaseven{\delta_{SE1b} = 2 \delta(0) D^{\alpha \rho \gamma}
D^{\beta \bar{\rho} \bar{\gamma}} {1 \over j_\gamma (j_\gamma + 2)}}
which cancels the first term in $\delta_{SE1f}$, while the second term
gives a finite contribution
\eqn\pthree{\eqalign{ G_{SE1b} & = -{6
\over \pi^2} ({1 \over \epsilon} + {1 \over 2} \ln(\pi) + {1 \over
2} \gamma) + {1 \over \pi^2} + {\cal O} (\epsilon). }}

For diagram SE1c, we find
\eqn\forpieight{\Pi_{SE1c} =  -4 C^{\rho \alpha  \sigma} C^{\bar{\sigma}
\beta \bar{\rho} }  \int {d \omega d^d k \over (2 \pi)^{d+1}} {1
\over (k^2 + j_\rho (j_\rho + 2))(k^2 + j_\sigma (j_\sigma + 2))}.}
This gives only a $\delta(0)$ contribution,
\eqn\fordeltaeight{\delta_{SE1c} = -4 \delta(0) C^{\rho \alpha  \sigma}
C^{\bar{\sigma} \beta \bar{\rho} }   {1 \over j_\rho (j_\rho + 2)
j_\sigma (j_\sigma + 2)}}
with no remainder,
\eqn\forreight{G_{SE1c} = 0.}

The ghost loop diagram SE1d also gives only a $\delta(0)$ term
\eqn\fordeltanine{\delta_{SE1d} = 2 \delta(0) C^{\rho \alpha  \sigma}
C^{\bar{\sigma} \beta \bar{\rho} }   {1 \over j_\rho (j_\rho + 2)
j_\sigma (j_\sigma + 2)}}
which, together with $\delta_{SE1c}$, cancels the second term in $\delta_{SE1f}$.

The remaining diagrams contain internal ${\cal A}$ lines, so for each
of these, it is only necessary to isolate the logarithmically
divergent term in the sum. From diagram SE1i, we find
\eqn\forpiten{\Pi_{SE1i} = - 2 {C^{\sigma \alpha \lambda}
C^{\bar{\sigma} \beta \bar{\lambda}} \over j_\lambda (j_\lambda +
2)} \int {d \omega d^d k \over (2 \pi)^{d+1}} {k^2 \over (k^2 +
j_\sigma (j_\sigma + 2))(\omega^2 + k^2 + j_\lambda (j_\lambda +
2))},}
where we have subtracted a term which gives a $\delta(0)$ from the
$\omega$ integral because it is multiplied by a $k$ integral which
vanishes for $\epsilon \to 0$. This gives a finite term
\eqn\forrten{G_{SE1i} = {1 \over 2 \pi^2}.}

For SE1h, we have the contribution
\eqn\piten{\eqalign{ \Pi_{SE1h} &= 2 D^{\alpha \rho \lambda} D^{\beta
\bar{\rho} \bar{\lambda}}(1 - {(j_\rho + 1)^2 - (j_\alpha + 1)^2
\over j_\lambda (j_\lambda + 2)})(1 - {(j_\rho + 1)^2 - (j_\beta +
1)^2 \over j_\lambda (j_\lambda + 2)})\cr & \cdot j_\lambda
(j_\lambda + 2) \int {d \omega d^d k \over (2 \pi)^{d+1}} {k^2 \over
(\omega^2 + k^2 + (j_\rho + 1)^2)(\omega^2 + k^2 + j_\lambda
(j_\lambda + 2))}{1 \over k^2 + j_\lambda (j_\lambda + 2)}. }}
This gives a finite contribution
\eqn\forreleven{G_{SE1h} = -{4 \over 3 \pi^2} \; .}

Finally, we find that diagram SE1j gives a net
contribution of
\eqn\pieleven{\eqalign{ \Pi_{SE1j} = -C^{\rho \alpha
\sigma} C^{\bar{\sigma} \beta \bar{\rho}} & \int {d \omega d^d k
\over (2 \pi)^{d+1}} {1 \over (\omega^2 + k^2 + j_\sigma (j_\sigma +
2))(\omega^2 + k^2 + j_\rho (j_\rho + 2))} \cr &\cdot \left\{ 4d + 2
k^2 (B_\alpha + B_\beta - 2 (E_\rho + E_\sigma)) \right.\cr & \left.
+k^4((B_\alpha - 2 E_\sigma)(B_\beta - 2 E_\rho) - 2 B_\alpha
E_\sigma - 2 B_\beta E_\rho) \right. \cr &\left.  - k^6((B_\beta - 2
E_\rho) B_\alpha E_\sigma + (B_\alpha - 2 E_\sigma) B_\beta E_\rho)
+ k^8 B_\alpha B_\beta E_\rho E_\sigma \right\}, }}
where we have defined
\eqn\erho{E_\rho \equiv {1 \over k^2 + j_\rho (j_\rho + 2)}, \qquad
B_\alpha \equiv {j_\sigma(j_\sigma + 2) + j_\rho (j_\rho + 2) + (j_\alpha
+ 1)^2 \over j_\sigma(j_\sigma + 2) j_\rho (j_\rho + 2)}.}
In evaluating this, things simplify, since we take $j_\alpha =
j_\beta = 1$ and this forces $j_\rho = j_\sigma$.
We have, as an intermediate step,
\eqn\inter{\eqalign{
G_{SE1j} &= \sum_a {a(a+1)^2(a+2) \over \pi^2}
\int {d^d k \over (2 \pi)^d} {1 \over (k^2 + a(a+2))^{3 \over 2}}
\left\{(d-1) + \left( {(1 + {1 \over 2} B k^2)a(a+2) \over k^2 + a(a+2)}
\right)^2 \right\},
}}
where
\eqn\bdef{B \equiv 2 {a(a+2) + 2 \over a^2(a+2)^2}.}
After computing the integrals, we find all terms are proportional to
$d$. Expanding in powers of $1/a$ and keeping only $1/a^{1-d}$ terms in
the summand, we find a net contribution of
\eqn\forrtwelve{G_{SE1j} = -{11 \over 6 \pi^2}.}

Combining all terms, our total result is
\eqn\forrdimreg{G_{dimreg} = {6 \over \pi^2} \left[{1 \over
\epsilon} - {1 \over 2} \gamma + {1 \over 2} \ln(\pi) + 1 \right] +
{6 \over \pi^2} \gamma + {3 \over 2 \pi^2} - {10 \over \pi^2}
\zeta(3).}
Using the modified minimal subtraction scheme defined in \S3.1, the
quantity in square brackets is set to zero, so we find finally that
\eqn\rdimreg{ G_{dimreg} = {6 \over \pi^2} \gamma + {3 \over 2
\pi^2} - {10 \over \pi^2} \zeta(3). }

\vskip 0.1 in
\noindent
{\bf Cutoffs and Counterterms}
\vskip 0.1 in

We now repeat the calculation using a regulator function
$R(\sqrt{-\nabla^2/M^2})$. Our scheme applies this only to the $A_i$
lines, for which $-\nabla^2$ gives $(j+1)^2$ for the mode with total
angular momentum quantum number $j$. The expressions $\Pi_i$ for the
diagrams are the same as above, so we simply set $d=0$ and insert
the regulator functions. We express our results in terms of the
basic integrals defined in \regdefs, using the Euler-McLaurin formula
to compare infinite sums with integrals.

For diagram SE1e, we find
%
\eqn\ccon{G_{SE1e}' = -
{4 \over \pi^2} \sum_{a=1}^\infty a R({a \over M}) + {4 \over \pi^2}
\sum_{a=1}^\infty {1 \over a} R({a \over M}) = -16 M^2 C_1 + {4
\over \pi^2} \ln \left( {{\cal A}_1 M \over \mu} \right) + {1 \over
3 \pi^2} + {4 \gamma \over \pi^2}. }
For diagram SE1a, we find
\eqn\cctw{\eqalign{ G_{SE1a}' &= {2 \over \pi^2}
\sum_{a=1}^\infty a R^2({a \over M}) + {8 \over \pi^2}
\sum_{a=1}^\infty {1 \over a} R^2({a \over M}) - {10 \over \pi^2}
\sum_{a=1}^\infty {1 \over a^3} \cr &= 8 M^2 C_2 + {8 \over \pi^2}
\ln \left( {{\cal A}_2 M \over \mu} \right) - {1 \over 6 \pi^2} + {8
\gamma \over \pi^2} - {10 \over \pi^2} \zeta(3) \; . }}
Diagram SE1b contains a $\delta(0)$ piece that is cancelled by the remaining diagrams.  This leaves the result
\eqn\ccth{\eqalign{ G_{SE1b}'
=& -{1 \over \pi^2} \sum_{a=3}^\infty {a^2-1 \over a-2} R({a \over
M}) -{1 \over \pi^2} \sum_{a=2}^\infty {a^2-1 \over a+2} R({a \over
M}) \cr
 =& - {3 \over 4 \pi^2} -{2 \over \pi^2} \sum_{a=3}^\infty
{a(a^2-1) \over (a^2-4)} R({a \over M}) \cr =& \, {23 \over
{2\pi^2}} - {2 \over \pi^2} \sum_{a=1}^\infty a R({a \over M}) - {6
\over \pi^2} \sum_{a=1}^\infty {1 \over a} R({a \over M}) \cr =& -8
M^2 C_1 - {6 \over \pi^2} \ln \left( {{\cal A}_1 M \over \mu}
\right) + {35 \over 3 \pi^2} - {6 \gamma \over \pi^2}. }}
Combining all terms so far, we have
\eqn\forroneloop{ G_{1-loop} = -24 M^2 C_1 + 8 M^2 C_2  - {2 \over
\pi^2} \ln \left( {{\cal A}_1 M \over \mu} \right) + {8 \over \pi^2}
\ln \left( {{\cal A}_2 M \over \mu} \right) + {71 \over 6 \pi^2}
-{10 \over \pi^2} \zeta(3) + {6\gamma \over \pi^2}.}

\fig{Counterterm contribution to $\langle A_iA_j\rangle$.}
{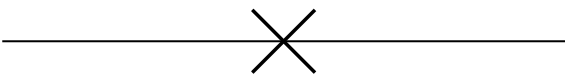}{2.0truein} \figlabel{\twopointctfig}

We have in addition the counterterm diagram of figure \twopointctfig,
which receives contributions from the $\tr(A_i A_i)$ and $p^2 \tr(A_i
A_i)$ counterterms. These give:
\eqn\forrct{G_{ct} = -12 Z_0 - 48 Z_1 = - 12 Z_0 - {6\over \pi^2}
\ln({M\over \mu}) - {16\over 5} F_2 - {7\over \pi^2} + {6\over
5\pi^2} \ln({{\cal A}_1^4 \over {\cal A}_2^9}).}
The final result is
\eqn\forrcutoff{G_{cutoff} = G_{1-loop} + G_{ct} =  -12 Z_0  -24
M^2 C_1 + 8 M^2 C_2 - {16 \over 5} F_2 + {14 \over 5\pi^2}
\ln({{\cal A}_1 \over {\cal A}_2}) -{10 \over \pi^2} \zeta(3)+ {29
\over 6 \pi^2} + {6\gamma \over \pi^2}.}

Demanding that this equals the dimensionally regularized result
\rdimreg\ above then determines the coefficient $Z_0$ to be
\eqn\curv{ Z_0 = -{4 \over 15} F_2 +{7 \over 30 \pi^2} \ln({{\cal
A}_1\over {\cal A}_2}) +{5 \over 18 \pi^2} - 2 M^2 C_1 + {2 \over 3}
M^2 C_2 .}

Thus, the required counterterms for our two-loop calculation are
given in \cterm, with the coefficients $Z_0$, $Z_1$, and $Z_2$ given
in \curv\ and \cts, respectively.

\newsec{Evaluating the two-loop diagrams}

\fig{Diagrams that contribute to the 2-loop free energy.  
The vertex of diagram 2a (2c) corresponds to the first (second) 
interaction term of \quartic.}{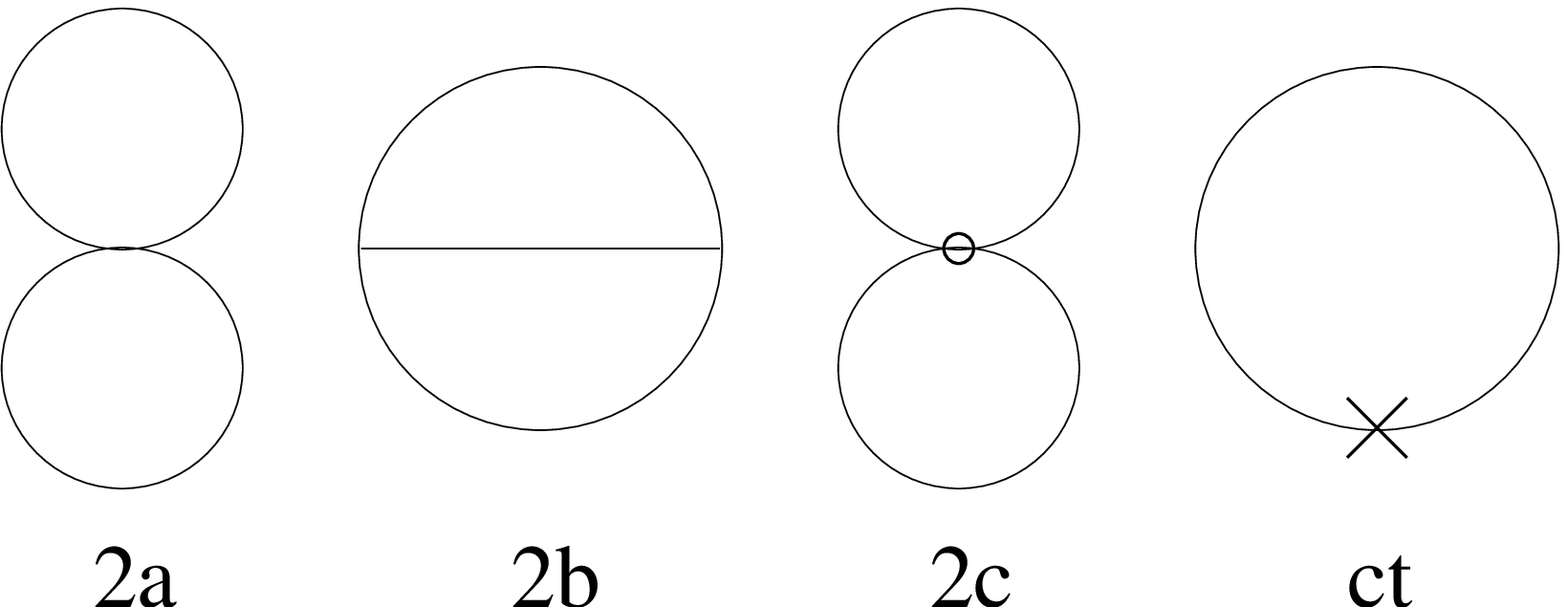}{4.0truein} \figlabel{\fediags}

Now that we have worked out the required counterterms, we are ready to
calculate the necessary correlators in \sefftwo\ to determine the
partition function. The relevant diagrams are depicted in figure
\fediags.  We begin first with the planar contributions.

\subsec{Two-loop planar diagrams}

From the discussion in section 2.4, the contribution to the
partition function from planar diagrams comes from the coefficients
of $\tr(U^n) \tr(U^\dagger {}^n)$ in the expansion in powers of $U$.
We have three two-loop planar diagrams, one from a pair of cubic
vertices and one from each of the two quartic vertices. In addition,
we have at the same order
one-loop planar diagrams for each of the counterterms \cterm.

We begin with the two-loop diagrams, including
a regulator function for each $A_i$ line, though we can neglect
regulator functions for momenta whose sums are already exponentially
suppressed. Full expressions for these diagrams were previously
derived in \firstord. They are\foot{Recall that in diagram 2c, we
ignore the part proportional to $\delta(0)$.} (with summation over
the spherical harmonic indices $\alpha$, $\beta$ and $\gamma$
implied)
\eqn\stwoa{\eqalign{ S^{pl}_{2a} &= - \beta {g_{YM}^2 \over 2}
(D^{\alpha \beta  \gamma} D^{\bar{\alpha} \bar{\beta} \bar{\gamma}}
- D^{\alpha \bar{\alpha} \gamma} D^{\beta \bar{\beta} \bar{\gamma}})
\Delta_{j_\alpha}(0, \alpha_{ab})
 \Delta_{j_\beta}(0, \alpha_{ac})  \cr
&={2 \beta g_{YM}^2 \over 3 \pi^2}  j_\alpha (j_\alpha + 2)
 j_\beta (j_\beta + 2)  \Delta_{j_\alpha}(0, \alpha_{ab})
 \Delta_{j_\beta}(0, \alpha_{ac}), }}
\eqn\stwob{\eqalign{
S^{pl}_{2b} &= - {\beta g_{YM}^2 \over 6}
\hat{E}^{\alpha \beta \gamma} \hat{E}^{\bar{\alpha}
\bar{\beta}\bar{\gamma}} \int dt \Delta_{j_\alpha}(t, \alpha_{ab})
\Delta_{j_\beta}(t, \alpha_{bc}) \Delta_{j_\gamma}(t, \alpha_{ca})
\cr &= - \beta g_{YM}^2 ({1 \over 3} (j_\alpha + j_\beta + j_\gamma
+ 3)^2 R_{4+}^2(j_\alpha,j_\beta,j_\gamma) + (j_\alpha + j_\beta -
j_\gamma + 1)^2 R_{4-}^2(j_\alpha,j_\beta,j_\gamma)) \cr & \qquad
\int dt \Delta_{j_\alpha}(t, \alpha_{ab}) \Delta_{j_\beta}(t,
\alpha_{bc}) \Delta_{j_\gamma}(t, \alpha_{ca}), }}
\eqn\stwoc{\eqalign{ S^{pl}_{2c} &=
\beta g_{YM}^2 {D^{\alpha \beta \gamma} D^{\bar{\alpha} \bar{\beta}
\bar{\gamma}} \over j_\gamma (j_\gamma+2)} (D_\tau
\Delta_{j_\alpha}(0, \alpha_{ac}) D_\tau \Delta_{j_\beta}(0,
\alpha_{ab}) + (j_\beta + 1)^2 \Delta_{j_\alpha}(0, \alpha_{bc})
\Delta_{j_\beta}(0, \alpha_{ab}) ) \cr &={2 \beta g_{YM}^2 \over
j_\gamma (j_\gamma+2)} (R_{3+}^2(j_\alpha,j_\gamma,j_\beta) +
R_{3-}^2(j_\alpha,j_\gamma,j_\beta)) \cr & \qquad \qquad (D_\tau
\Delta_{j_\beta} (0, \alpha_{ab}) D_\tau \Delta_{j_\alpha} (0,
\alpha_{ac}) + (j_\beta + 1)^2 \Delta_{j_\alpha} (0, \alpha_{bc})
\Delta_{j_\beta}(0, \alpha_{ab})), }}
where $\hat{E}$ and the functions $R_{3 \pm}$ and $R_{4 \pm}$ are
defined in appendix B. Here, the first line for each diagram gives
the expression for the diagram before performing any sums over
angular momenta, while the second line gives the result after
summing over everything but total angular momentum quantum numbers
(these come into the regulator, which we have not yet written
explicitly). Note that for all diagrams, each of the propagators
contributes factors of $\alpha$ to two of the three index loops,
which we label by $a$, $b$, and $c$. The notation $\alpha_{ab}$
indicates that for the tensor products $(\alpha \otimes 1 - 1
\otimes \alpha)$ appearing in the propagator \defdelta, the first and second
elements of the tensor product appear in the traces associated with
index loops $a$ and $b$, respectively.

We now expand these expressions in powers of $U$ to read off the
required coefficients $f_n(x)$ defined in \planars, inserting
explicit expressions for the regulators at this stage. For all
diagrams, we find that
\eqn\forfn{ f_n(x) = f_1(x^n),}
so we need only give the results for $f_1(x)$.

We find that diagram 2a gives
\eqn\ftwoa{ f^{2a}_1(x) = {1 \over 3 \pi^2} (f(x))^2 + {2
\over 3 \pi^2} f(x) \left\{ \sum_{a=1}^\infty a R({a \over M}) -
\sum_{a=1}^\infty {1 \over a} R({a \over M}) \right\}, }
where
\eqn\forfx{f(x) \equiv \sum_{b=1}^\infty {b(b+2) \over (b+1)}
x^{b+1} = {x \over (1-x)^2} + \ln(1-x).}

For diagram 2b, we find
\eqn\ftwob{\eqalign{ f^{2b}_1(x) &=
\sum_{a=1}^\infty \sum_{b=1}^\infty
\sum_{c/2=(|a-b|+1)/2}^{(a+b-1)/2} {(b+1)x^{a+c+2} - (a+c+2)x^{b+1}
\over 2(a+1)(b+1)(c+1)(a+b+c+3)(a+c-b+1) } \cr & \cdot \left\{
R_{4+}^2(a,b,c)(a+b+c+3)^2 + R_{4-}^2(a,b;c)(a+b-c+1)^2 \right. \cr
& +  \left. R_{4-}^2(b,c;a)(b+c-a+1)^2 +
R_{4-}^2(c,a;b)(c+a-b+1)^2\right\} R({a+1 \over M}) R({c+1 \over
M}). }}
To see the regulator dependent pieces explicitly, we may
expand $R({c+1 \over M})$ about $c=a$ for large $a$,
\eqn\forrc{R({c+1 \over M}) = R({a+1 \over M}) + {c-a \over
M}R'({a+1 \over M}) + {(c-a)^2 \over 2 M^2} R''({a+1 \over M}) +
\dots}
For large $a$, the sum over $c$ of the summand above with various powers of
$(c-a)$
inserted then has the following asymptotic behavior as a function of $a$:
\eqn\sumcs{\eqalign{ \sum &\rightarrow - {1 \over 3 \pi^2} a {b^2-1
\over b} x^b - {1 \over 15 \pi^2} {1 \over a} {(b^2-1)(8b^2-7) \over
b} x^b + {\cal O}(1/a^2), \cr \sum (c-a) &\rightarrow -{1 \over 30
\pi^2} {(b^2-1)(b^2-4) \over b} x^b +  {\cal O}(1/a), \cr \sum
(c-a)^2 &\rightarrow -{1 \over 15 \pi^2} a {(b^2-1)(b^2-4) \over b}
x^b + {\cal O}(1), \cr }}
where we have made the replacements $a \to
a-1$, $b \to b-1$. The omitted terms lead only to
regulator-independent finite contributions. The regulator-dependent
terms are then 
\eqn\fregb{\eqalign{ f^{reg}_{2b} =& - {1 \over 3
\pi^2} \sum_{b=1}^\infty {b^2 -1 \over b} x^b \sum_{a=1}^\infty a
R^2({a \over M}) \cr & - {1 \over 15 \pi^2} \sum_{b=1}^\infty {(b^2
-1) (8b^2-7) \over b} x^b \sum_{a=1}^\infty {1 \over a} R^2({a \over
M})  \cr &-{1 \over M^2} {1 \over 30 \pi^2} \sum_{b=1}^\infty {(b^2
-1) (b^2-4) \over b} x^b \sum_{a=1}^\infty a R({a \over M}) R''({a
\over M}). }} 
Note that the term involving $1/M \sum_a R(a/M) R'(a/M)$
is regulator-independent.

For diagram 2c, we find
\eqn\ftwoc{\eqalign{ f^{2c}_1 &= \sum_{a=1}^\infty \sum_{b=1}^\infty
\sum_{c/2=(||a-b|-1|+1)/2}^{(a+b)/2} {1 \over c(c+2)} \left\{{b-a
\over a+1} x^{a+b+2} + \left({b+1 \over a+1} + {a+1 \over b+1}
\right) x^{b+1} \right\}  \cr & \qquad \qquad \qquad \qquad \cdot
\left\{ R_{3+}^2(a,c,b)+ R_{3-}^2(a,c,b) \right\} R({a+1 \over M}).
}}
The asymptotic
behavior of the sum over $c$ for large $a$ is \eqn\sumcsc{ \sum
\rightarrow  {1 \over 3 \pi^2} a {b^2-1 \over b} x^b + {1 \over 15
\pi^2} {1 \over a} {(b^2-1)(8b^2+3) \over b} x^b + {\cal O}(1/a^2).}
The regulator-dependent terms are then
\eqn\fregb{\eqalign{
f^{reg}_{2c} =& {1 \over 3 \pi^2} \sum_{b=1}^\infty {b^2 -1 \over b}
x^b \sum_{a=1}^\infty a R({a \over M}) \cr & + {1 \over 15 \pi^2}
\sum_{b=1}^\infty {(b^2 -1) (8b^2+3) \over b} x^b \sum_{a=1}^\infty {1
\over a} R({a \over M}).  \cr }}
Combining all the regulator-dependent terms 
and carefully comparing the divergent sums to the
divergent integrals defined in \regdefs, we find
\eqn\fregtot{ f^{reg} = f(x) (4 C_1 M^2  - {4\over 3} C_2
M^2 -{1 \over 18 \pi^2}) + {1\over {15 \pi^2}}
(8 g(x) - 7 f(x)) \ln({{\cal
A}_1 \over {\cal A}_2}) + {2 \over 15} F_2 (4 f(x)-g(x)), }
where $f(x)$ is as above, and
\eqn\forgx{g(x) \equiv \sum_{b=1}^\infty b(b^2-1) x^b = {6x^2 \over
(1-x)^4} \; .}
The expression \fregtot\ will be useful in verifying that all
regulator dependence cancels, though in practice, we simply need to
use \ftwoa, \ftwob, and \ftwoc\ with any convenient choice of
regulator.

\subsec{Counterterms}

To the expressions above, we must add the counterterm contributions,
coming from the single-trace counterterms in \cterm. Using this
expression, we find
\eqn\forseff{S_{eff}^{ct} = \langle S_{ct} \rangle = \beta g_{YM}^2
N \sum_{\alpha} (Z_0 + (Z_1 - Z_2) (j_\alpha + 1)^2)
\Delta_{j_{\alpha}}(0, \alpha_{ab}).}
To determine the term proportional to $Z_2$, we have used the fact
that
\eqn\propdef{ (-\tdzero^2 +(j+1)^2 )\Delta_j(t,\alpha)= \delta(t) \; ,}
ignoring the resulting $\delta(0)$ term, which will be cancelled by
another diagram which is independent of $U$ and the temperature.

Expanding this to find the coefficient of $\tr(U^n) \tr(U^\dagger
{}^n)$ and performing the angular momentum sum, we find that the
counterterm contribution to $f_n$ is $f^{ct}_n(x)=f^{ct}_1(x^n)$,
where
\eqn\forfone{f^{ct}_1(x) = 2 Z_0 f(x) + 2 (Z_1 - Z_2)g(x).}
Using the expressions \cts\ and \curv\ above, this becomes
\eqn\counter{\eqalign{ f^{ct}_1(x) =& f(x) \left[ M^2 (-4 C_1 + {4\over 3} C_2) +
{7 \over {15\pi^2}} \ln\left({{\cal A}_1\over {\cal A}_2}\right)
- {8 \over 15} F_2 + {5 \over 9 \pi^2}\right] \cr &+ g(x) \left[{2 \over 15}
F_2 + {1\over 4\pi^2} - {8 \over {15\pi^2}} \ln\left({{\cal A}_1\over {\cal A}_2}\right)
\right]. }}
Comparing this with the regulator-dependent pieces \fregtot\ above, we see that all
regulator dependence cancels in the sum $f_1^{2a}+f_1^{2b}+f_1^{2c}+
f_1^{ct}$. Thus, our final result is
\eqn\nforfn{f_n(x) = f_1(x^n), }
where $f_1(x)$ is the sum of \counter, \ftwoa, \ftwob, and \ftwoc,
which is finite and regulator independent. This may be computed
numerically for any value of $x$, and we do indeed find that the
numerical results are independent of the regulator used. The
perturbative expansion of $f_1$ at small $x$ takes the form
\eqn\expfone{f_1(x) = {1\over 4\pi^2} x^2 + {4\over \pi^2} x^3 +
{55\over 4\pi^2} x^4 + \cdots,}
and its value at the critical temperature \hagtemp\ is given by
\eqn\fonehag{f_1(x_{c,0}) \simeq 0.0253.}
The functions $f_n(x)$ determine the planar contribution to the
two-loop partition function via \twoform\ while $f_1(x)$ determines
the perturbative shift in the Hagedorn temperature, which we compute
using \xshift\ in section 4.5 below.

\subsec{Two-loop non-planar diagrams}

We now turn to the non-planar contribution to the two-loop partition
function. In this case, since there is only a single index loop, and
since each term in the propagators contributes an equal number of
$U$'s and $U^\dagger$'s, we will always end up with just the
identity matrix inside the single trace.  Equivalently, we can
simply set $\alpha=0$ ($U=1$) in all propagators from the start. The
resulting temperature-dependent but $U$-independent expressions
contribute directly to the two-loop free energy via \fcorr.

We find that the expressions for the three non-planar diagrams are
related to the expressions \stwoa-\stwoc\ 
for the planar diagrams by setting
$\alpha=0$ and including an overall factor of $-1/N^2$.\foot{This is
related to the fact that the planar and non-planar diagrams must
cancel for the Abelian theory with $N=1$, which is free.} Thus, we
have
\eqn\stwonpa{\eqalign{ S^{np}_{2a} &=  \beta {g_{YM}^2 \over 2}
(D^{\alpha \beta  \gamma} D^{\bar{\alpha} \bar{\beta} \bar{\gamma}}
- D^{\alpha \bar{\alpha} \gamma} D^{\beta \bar{\beta} \bar{\gamma}})
\Delta_{j_\alpha}(0, 0)
 \Delta_{j_\beta}(0, 0)  \cr
&=-{2 \beta g_{YM}^2 \over 3 \pi^2}  j_\alpha (j_\alpha + 2)
 j_\beta (j_\beta + 2)  \Delta_{j_\alpha}(0, 0)
 \Delta_{j_\beta}(0, 0), }}
\eqn\stwonpb{\eqalign{
S^{np}_{2b} &=  {\beta g_{YM}^2 \over 6}
\hat{E}^{\alpha \beta \gamma} \hat{E}^{\bar{\alpha}
\bar{\beta}\bar{\gamma}} \int dt \Delta_{j_\alpha}(t, 0)
\Delta_{j_\beta}(t, 0) \Delta_{j_\gamma}(t, 0) \cr &=  \beta
g_{YM}^2 \left({1 \over 3} (j_\alpha + j_\beta + j_\gamma + 3)^2
R_{4+}^2(j_\alpha,j_\beta,j_\gamma) + (j_\alpha + j_\beta - j_\gamma
+ 1)^2 R_{4-}^2(j_\alpha,j_\beta,j_\gamma)\right) \cr & \qquad \int dt
\Delta_{j_\alpha}(t, 0) \Delta_{j_\beta}(t, 0) \Delta_{j_\gamma}(t,
0), }}
\eqn\stwonpc{\eqalign{S^{np}_{2c} &= -\beta g_{YM}^2 {D^{\alpha \beta \gamma}
D^{\bar{\alpha} \bar{\beta} \bar{\gamma}} \over j_\gamma
(j_\gamma+2)} (D_\tau \Delta_{j_\alpha}(0, 0) D_\tau
\Delta_{j_\beta}(0, 0) + (j_\beta + 1)^2 \Delta_{j_\alpha}(0, 0)
\Delta_{j_\beta}(0, 0) ) \cr &=-{2 \beta g_{YM}^2 \over j_\gamma
(j_\gamma+2)} (R_{3+}^2(j_\alpha,j_\gamma,j_\beta) +
R_{3-}^2(j_\alpha,j_\gamma,j_\beta)) \cr & \qquad \qquad (D_\tau
\Delta_{j_\beta} (0, 0) D_\tau \Delta_{j_\alpha} (0, 0) + (j_\beta +
1)^2 \Delta_{j_\alpha} (0, 0) \Delta_{j_\beta}(0, 0)). }}
In each case, the group indices are contracted into a single
(trivial) trace which gives an overall factor of $N$. We will find
again that the angular momentum sums here are divergent, but there
are additional counterterms coming from the double-trace terms in
\cterm. The result for these is obtained from the result for the
single-trace counterterms in the same way that the non-planar
two-loop contributions are obtained from the planar two-loop
contributions.

Using this relation between planar diagrams/single-trace
counterterms and non-planar diagrams/double-trace counterterms, it
follows from \planars\ that the full contribution to 
$S^{\rm eff}_{\rm 2-loop}$ may be written as
\eqn\plnp{\eqalign{ S_{pl+np}^{\rm eff} &= \beta g_{YM}^2 N
\sum_{n=1}^{\infty}
f_n(x)(\tr(U^n) \tr(U^\dagger {}^n) - 1) \cr & \qquad + \beta
g_{YM}^2 \sum_{n \ge m > 0} f_{nm}(x) \{ \tr(U^n) \tr(U^m)
\tr(U^{-n-m}) + c.c. - 2N\} \; . }}
For the planar diagrams, the
two-trace terms in the first line arose from the combination of
divergent two-loop contributions plus counterterms to yield the
finite regulator-independent results defined by $f_n(x) = f_1(x^n)$.
While we did not calculate the three trace terms explicitly, it is
straightforward to show that they are manifestly finite. It
immediately follows that all divergences and regulator-dependence
cancel also for the non-planar diagrams plus double-trace
counterterms.

Focusing back on the non-planar terms that we need to calculate, it
is convenient to use \plnp\ to write
\eqn\fdecomp{ F_2^{np}(x) = -\sum_{n=1}^\infty f_1(x^n) +
\tilde{F}^{np}(x) \; , }
since we have
already computed $f_1(x)$; it remains only to calculate
\eqn\forftilde{\tilde{F}^{np}_2(x) = -2 \sum_{n \ge m > 0} f_{nm}(x)
\; ,}
where $f_{nm}(x)$ may be computed from the expressions \stwoa-\stwoc.
Diagram by diagram, we find
\eqn\ftwoanp{\eqalign{ \tilde{F}^{np}_{2a} &  = -
\sum_{\ja=0}^\infty \sum_{\jb=0}^\infty {1 \over {6 \pi^2}}
{\ja(\ja+2) \jb(\jb+2) \over {(\ja+1)(\jb+1)}} \cr & \left({(1 +
x^{\ja+1}) (1 + x^{\jb+1}) \over (1 - x^{\ja+1}) (1 - x^{\jb+1})} -
{(1 + x^{\jb+1}) \over (1 - x^{\jb+1})} - {(1 + x^{\ja+1}) \over (1
- x^{\ja+1})} - 2{x^{\ja+\jb+2} \over 1-x^{\ja+\jb+2}} + 1 \right),
\cr}}
\eqn\ftwobnp{\eqalign{ \tilde{F}^{np}_{2b} &= \sum_{\ja=1}^\infty
\sum_{\jb=1}^\infty \sum_{\jc/2=(|\ja-\jb|+1)/2}^{(\ja+\jb-1)/2} \cr
&\qquad \qquad  {(R_{4+}^2(\ja,\jb,\jc)(\ja+\jb+\jc+3)^2 + 3
R_{4-}^2(\ja,\jb;\jc)(\ja+\jb-\jc+1)^2 ) \over 12
(\ja+1)(\jb+1)(\jc+1)} \cr & \left[ {1 \over \ja+\jb+\jc+3} - {1
\over (1-x^{\ja+1})(1-x^{\jb+1})(1-x^{\jc+1})} \cdot \right. \cr & \qquad
\left\{ {1-x^{\ja+\jb+\jc+3} \over
\ja+\jb+\jc+3}+{x^{\ja+1}-x^{\jb+\jc+2} \over
\jb+\jc-\ja+1}+{x^{\jb+1}-x^{\ja+\jc+2} \over
\ja+\jc-\jb+1}+{x^{\jc+1}-x^{\ja+\jb+2} \over \ja+\jb-\jc+1}
\right\} \cr &+{2 \over (\ja+\jb+\jc+3)(\ja+\jb-\jc+1)}
\left\{{(\ja+\jb+1)x^{\jc+1} \over 1-x^{\jc+1}} -{(\jc+1)
x^{\ja+\jb+2} \over 1-x^{\ja+\jb+2}}  \right\}  \cr &+{2 \over
(\ja+\jb+\jc+3)(\jb+\jc-\ja+1)} \left\{{(\jb+\jc+1)x^{\ja+1} \over
1-x^{\ja+1}} -{(\ja+1) x^{\jb+\jc+2} \over 1-x^{\jb+\jc+2}} \right\}
\cr &\left. +{2 \over (\ja+\jb+\jc+3)(\jc+\ja-\jb+1)}
\left\{{(\jc+\ja+1)x^{\jb+1} \over 1-x^{\jb+1}} -{(\jb+1)
x^{\jc+\ja+2} \over 1-x^{\jc+\ja+2}}  \right\}  \right], }}
and
\eqn\ftwocnp{\eqalign{ \tilde{F}^{np}_{2c} & = \sum_{\ja=1}^\infty
\sum_{\jb=1}^\infty \sum_{\jc/2=(||\ja-\jb|-1|+1)/2}^{(\ja+\jb)/2}
{1 \over \jc(\jc+2)} (R_{3+}^2(\ja,\jc,\jb)+ R_{3-}^2(\ja,\jc,\jb) )
\cr & \left\{ {\jb+1 \over \ja+1} \left({(1 + x^{\ja+1}) (1 +
x^{\jb+1}) \over (1 - x^{\ja+1}) (1 - x^{\jb+1})} - {(1 + x^{\jb+1})
\over (1 - x^{\jb+1})} - {(1 + x^{\ja+1}) \over (1 - x^{\ja+1})}
\right) \right. \cr &\qquad \qquad \qquad \left. - \left({\jb+1
\over \ja+1} - 1 \right){2 x^{\ja+\jb+2} \over (1-x^{\ja+\jb+2})}
\right\}. }}
Using \fdecomp, these results, combined with our results for
$f_1(x)$ in section 4.2, give the final result for $F_2^{np}$, which
in turn gives the non-planar contribution to the two-loop partition
function and free energy via \twoform\ and \fcorr.

\subsec{Results: Two-loop partition function}

We have now calculated all the elements necessary to give the full
two-loop partition function. Using \twoform\ and \fdecomp, we find
that the two-loop partition function for $U(N)$ Yang-Mills theory at
large $N$ on a small $S^3$ is
\eqn\partun{\eqalign{ Z_{2-loop}^{U(N)} &=  
e^{- \lambda \beta \tilde{F}^{np}_2(x)}
\prod_{n=1}^\infty {e^{\lambda \beta f_1(x^n)} \over 1 - z(x^n) +
\lambda n \beta f_1(x^n)} + {\cal O} (\lambda^2) \; , }}
where
$z(x)$ is given in \zdef, $\tilde{F}_2^{np}(x)$ is the sum of
\ftwoanp, \ftwobnp, and \ftwocnp, and $f_1(x)$ is the sum of \ftwoa,
\ftwob, \ftwoc, and \counter. To get the $SU(N)$ result, we simply
divide by $Z_{U(1)}$, yielding
\eqn\partsun{\eqalign{ Z_{2-loop}^{SU(N)} &=  
e^{- \lambda \beta \tilde{F}^{np}_2(x)}
\prod_{n=1}^\infty {e^{- z(x^n)/n + \lambda \beta f_1(x^n)} \over 1
- z(x^n) + \lambda n \beta f_1(x^n)} + {\cal O} (\lambda^2) \; . }}
For either $U(N)$ or $SU(N)$, the order $\lambda$ correction to the
free energy (using \fcorr) is
\eqn\fordeltaf{\delta F = -T \delta \ln(Z) = \lambda \left[
\tilde{F}_2^{np}(x)+ \sum_{n=1}^\infty f_1(x^n)\left({ n \over 1-z(x^n)}
- 1\right) \right] \; .}
Our results for the two-loop partition function can be expanded in
powers of $x$ as in \result, from which we can read off the sum of
energy corrections for all states with a given energy in the free
theory, since
\eqn\forpowerexp{\sum_i x^{E_0 + \lambda \delta E_i} = x^{E_0}\left(1 +
\lambda \ln(x) \sum_i \delta E_i + {\cal O}(\lambda^2) \right) \; .}
For $SU(N)$, we find that the first few terms in the series give
%
\eqn\zexp{Z^{SU(N)}_{2-loop} = 1 + \left(21 + {4\over \pi^2} \lambda
\ln(x)\right) x^4 + \left(96 + {28\over \pi^2} \lambda \ln(x)\right) x^5 + \left(392 +
{178\over \pi^2} \lambda \ln(x)\right) x^6 + \cdots. }
Thus, for example,
the sum of perturbative energy shifts for the 21 independent states
with energy $4/R_{S^3}$ is $4 \lambda / \pi^2$.

Since the free Yang-Mills theory in the small volume limit is
conformal, we have a map between states in this theory and local
operators in Euclidean Yang-Mills theory on $\IR^4$. The coefficient
of the term proportional to $x^n \ln(x)$ is then interpreted as the
sum of anomalous dimensions for all operators with dimension $n$.
Thus, we can use the results of \beisert\ for anomalous dimensions
of operators in pure Yang-Mills theory as a check on our results
(see section 5.1 below).

\subsec{Results: Order $\lambda$ correction to critical temperature}

As explained in section 2.4, the two-loop corrections to the
partition function will shift the critical temperature associated
with the Hagedorn and deconfinement transitions. To compute the
shift in $x_c$ or $T_c$, we use \xshift\ or \tshift, which only
involve the result \fonehag\ for $f_1(x_c)$.
Recalling from section 2.3 that $x_{c,0} = 2-\sqrt{3}$
for $\lambda =0$, we find
\eqn\fordeltaxc{\delta x_c =   0.00298 \lambda,}
so the transition occurs at
\eqn\fornewx{x_c = 2 - \sqrt{3} + 0.00298 \lambda + {\cal
O}(\lambda^2).}
Equivalently, from \tshift, the critical temperature to order
$\lambda$ is
\eqn\tsh{T_c R_{S^3} =  T_{c,0}R_{S^3} \cdot \left(1 +  
\frac{0.00298\lambda}{\left(2-\sqrt{3}\right)\ln\left(2+\sqrt{3}\right)} +
{\cal O} (\lambda^2) \right). }
Below, we will reproduce this result using
a formula in \spradlin\ for the shift in the transition temperature
from anomalous dimensions, providing a strong quantitative check of
our results.

\newsec{Checks}

\subsec{Anomalous dimensions for low-dimension operators}

As a first check on our results, we will try to reproduce our result
\zexp\ for the leading terms in the expansion of the $SU(N)$
partition function using the results of \beisert\ for anomalous
dimensions of operators in pure Yang-Mills theory.

Explicit results for the one-loop anomalous dimensions for
low-dimension operators in pure large $N$ $SU(N)$ Yang-Mills theory
are given in table 1 of \beisert. They show that two of the
dimension four states have anomalous dimension $-(11/3)
(\lambda/8\pi^2)$, ten have anomalous dimension $(7/3)
(\lambda/8\pi^2)$, and nine have vanishing anomalous dimension,
while at dimension five the $16$ primary states have anomalous
dimension $3 (\lambda / 8 \pi^2)$ (and each of the dimension four
primary states that has an anomalous dimension has four
descendants). Adding also all the dimension six states, we find that
the results of \beisert\ lead to
\eqn\forztwoloop{Z_{2-loop}(x) = 1 + \left(21 + {2\lambda \over \pi^2}
\ln(x)\right) x^4 + \left(96 + {14\lambda \over \pi^2} \ln(x)\right) x^5 + \left(392 + {89
\lambda \over \pi^2} \ln(x)\right) x^6 + \cdots.}
This precisely agrees with our result \zexp, if we recall that, due
to a difference in conventions, $\lambda$ in \beisert\ is twice the
$\lambda$ that we used in our computation (as was also the case for
the comparison of the flat-space two-loop partition functions in
\firstord).

\subsec{Shift in $T_c$ from anomalous dimensions}

In this section, we use an alternate method, based on \spradlin, to
check our result for the order $\lambda$ shift in the transition
temperature.

Single trace operators in pure Yang-Mills theory may be put in one
to one correspondence with spin chains of lengths
$l=1,2,\cdots,\infty$, where the spins in these chains are vectors
in one of two representations of the conformal group. The primaries
of these representations are the self-dual and anti-self-dual parts
of $F_{\mu \nu}$ and carry quantum numbers $(1,0, 2)$ and $(0, 1,
2)$, respectively, under $(j_1, j_2, \Delta)$ ($j_1$ labels the
representation of the first $SU(2)$, $j_2$ of the second $SU(2)$,
and $\Delta$ is the scaling dimension). We will call these two
representations chiral $c$ and anti-chiral ${\bar c}$, respectively.
It was demonstrated in \beisert\ that
\eqn\tensorprod{c \times c=
(0,0,4)+ (1,0,4)+ (2,0,4) +\sum_{j=1}^\infty (2+{j\over 2}, {j\over
2}, 4+j).}
The tensor product of ${\bar c}$ with ${\bar c}$ is
obtained from \tensorprod\ by a $j_1 \leftrightarrow j_2$ flip.
Finally,
\eqn\tenprod{c \times {\bar c}=\sum_{j=2}^\infty ({j \over
2}, {j \over 2}, 2+j).}

We now list $\chi_R=\Tr_{R} (x^\Delta)$, the
characters for all these representations (they will be needed below).
Actually, we will find it more convenient to list $\chi'_R\equiv (1-x)^4
\chi_R$ :
\eqn\characters{\eqalign{\chi'_{0,0, 4}&=x^4, ~~
\chi'_{0,1,4}=3x^4, ~~ \chi'_{0,2,4}= 5 x^4, \cr
\chi'_{2 +{j\over
2}, {j\over 2}, 4+j}&=(5+j)(j+1)x^{4+j} - j(4+j)x^{5+j},\cr
\chi'_{{j\over 2}, {j\over 2}, 2+j}&= (j+1)^2 x^{j+2}-j^2 x^{j+3}.}}

According to equation (6.10) of \spradlin,
\eqn\hagshift{ {\delta x_{c} \over x_{c,0}} =-{\lambda \ln (x_{c,0}) \over
4 \pi^2 x_{c,0} z'(x_{c,0})} \times \langle D_2(x_{c,0}) \rangle,}
where $z(x)$ was defined in \zdef,
$\vev{D_2(x)}$ is (see equations (4.4) and (4.8) in \beisert)\foot{
Note that here, unlike the previous subsection, we are using the
same conventions for $\lambda$ as in the rest of the paper.}
\eqn\Dt{\eqalign{\vev{D_2(x)}=&-{11\over 3} \chi_{0,0,4} +{1\over 3}
\chi_{0,1,4} + {7\over 3} \chi_{0,2,4} + \sum_{j=1}^\infty \left( 4
h(j+2)-{11\over 3} \right) \chi_{2 +{j\over 2}, {j\over 2}, 4+j}\cr
& + \sum_{j=2}^\infty 2( h(j-2) + h(j+2) - 11/6 ) \chi_{{j\over 2},
{j\over 2}, 2+j}, }}
and $h(j) \equiv \sum_{k=1}^j 1/k$ (with $h(0)=0$).

It is easy to numerically evaluate the sum \Dt\
at the Hagedorn temperature, or to compute it analytically using
\eqn\fordtwo{ \vev{D_2(x)} = {1 \over (1-x)^6} \left[ 2 (-\ln(1-x)) (1 -
3 x - 3 x^2 + x^3)^2
 - 2 x + 11 x^2 - {2\over 3} x^3 - 26 x^4 + 36 x^5 - {23\over 3} x^6 \right].
}
Using \hagshift, we find
\eqn\numev{\vev{D_2(x_{c,0})} = \left(28-16\sqrt{3}\right) / \left(\sqrt{3}-1\right)^4 = 1 \to
\delta T_c / T_{c,0} = {\lambda \over 12\pi^2} \langle D_2(x_{c,0}) \rangle =
{\lambda \over 12\pi^2},}
%
or
\eqn\predhagshift{\eqalign{
T_c R_{S^3} &= T_{c,0}R_{S^3} \cdot
\left(1 + \frac{\lambda}{12 \pi^2} + {\cal O}(\lambda^2)
\right) \cr
&= T_{c,0}R_{S^3} \cdot \left(1+ \frac{1}{\left(2-\sqrt{3}\right)
\ln\left(2+\sqrt{3}\right)}\left[\frac{\left(2-\sqrt{3}\right)
\ln\left(2+\sqrt{3}\right)\lambda}{12\pi^2} \right]
+ {\cal O}(\lambda^2) \right).
}}
Finally, given
\eqn\hagshiftcomp{\frac{\left(2-\sqrt{3}\right)\ln\left(2+\sqrt{3}\right)}{12\pi^2}\approx 0.00298 \; ,}
we find perfect agreement with \tsh\ (within
our numerical accuracy). Note that, in general, in order to have
agreement between \xshift\ and \hagshift, we should have
\eqn\agreement{f_1(x_{c,0}) = {{\vev{D_2(x_{c,0})}} \over 4\pi^2},}
which is indeed obeyed (within our accuracy) by our result \fonehag.

\bigskip

\centerline{\bf Acknowledgements}

We would like to thank Shiraz Minwalla and Kyriakos Papadodimas for
helpful discussions and collaboration during the early part of this
work. OA would like to thank Harvard University, the Aspen Center
for Physics and the Institute for Advanced Study for hospitality
during the course of this work. MVR would like to thank the Weizmann
Institute of Science for hospitality while part of this work was
completed. The work of OA was supported in part by the Israel-U.S.
Binational Science Foundation, by the Israel Science Foundation
(grant number 1399/04), by the Braun-Roger-Siegl foundation, by the
European network HPRN-CT-2000-00122, by a grant from the G.I.F., the
German-Israeli Foundation for Scientific Research and Development,
by Minerva, by a grant of DIP (H.52), and by the Schrum foundation. 
The work of JM was
supported in part by an NSF Graduate Research Fellowship and by 
DOE grant DE-FG01-91ER40654. The work
of MVR was supported in part by the Natural Sciences and Engineering
Council of Canada, by the Canada Research Chairs programme, and by
the Alfred P. Sloan Foundation.

\appendix{A}{Actions and propagators}

The quadratic action for gauge-fixed Euclidean pure Yang-Mills
theory on $S^3 \times S^1$, as described in \S2, is given by
(with $D_{\tau} \equiv \partial_t - i [\alpha, \cdot]$)
\eqn\quadr{\eqalign{ S_2 = \int dt \, \tr( {1 \over 2} A^{\bar{\alpha}}
( - \tdzero^2 + (j_\alpha+1)^2) A^\alpha + {1 \over 2}
a^{\bar{\alpha}} j_\alpha (j_\alpha + 2) a^\alpha +
\bar{c}^{\bar{\alpha}} j_\alpha (j_\alpha + 2) c^\alpha ). }}
In addition, we have cubic interactions
\eqn\cub{\eqalign{ S_3 = g_{YM} \int dt \, \tr( & i
\bar{c}^{\bar \alpha} [A^\gamma, c^\beta] C^{\bar{\alpha} \gamma
\beta} +2i a^\alpha A^\gamma a^\beta C^{\alpha \gamma \beta} \cr
&-i[A^\alpha, D_\tau A^\beta]a^\gamma D^{\alpha \beta \gamma} + i
 A^\alpha A^\beta A^\gamma \epsilon_\alpha (j_\alpha + 1)
E^{\alpha \beta \gamma}), }}
and quartic interactions
\eqn\quar{\eqalign{ S_4 = g_{YM}^2 \int dt \, \tr( & -{1 \over
2} [a^\alpha, A^\beta][a^\gamma, A^\delta]\left(D^{\beta
\bar{\lambda} \alpha} D^{\delta \lambda \gamma} + {1 \over
j_\lambda (j_\lambda+2)} C^{\alpha \beta \bar{\lambda}} C^{\gamma
\delta \lambda} \right) \cr & - {1 \over 2} A^\alpha A^\beta
A^\gamma A^\delta \left( D^{\alpha \gamma \bar{\lambda}} D^{\beta
\delta \lambda} - D^{\alpha \delta \bar{\lambda}} D^{\beta \gamma
\lambda} \right)).}}
The quantities $C$, $D$, and $E$ are integrals over spherical
harmonics, and are defined in appendix $B$. The propagators of the
various fields follow from \quadr\ and are given by
\eqn\propc{ \langle \bar{c}^{\bar \alpha}_{ab}(t) c_{cd}^\beta(t')
\rangle = {1 \over j_\alpha (j_\alpha +2)} \delta^{\alpha \beta}
\delta(t - t') \delta_{ad} \delta_{cb}, }
\eqn\propa{ \langle
a_{ab}^\alpha(t) a_{cd}^\beta(t') \rangle = {1 \over j_\alpha
(j_\alpha +2)} \delta^{\alpha \bar{\beta}} \delta(t - t')
\delta_{ad} \delta_{cb} , }
\eqn\propba{ \langle A^\alpha_{ab}(t)
A^\beta_{cd}(t') \rangle = \delta^{\alpha \bar{\beta}}
\Delta^{ad,cb}_{j_\alpha}(t-t',\alpha), }
where $\Delta$ is defined in section 2.

\appendix{B}{Spherical harmonics on $S^3$}

A detailed discussion of spherical harmonics on $S^3$ may be found
in appendix B of \firstord. Here, we collect various formulae
relevant for the present calculation. Many of the basic results were
derived in \Cutkosky.

\subsec{Basic properties of spherical harmonics}

Scalar functions on the sphere may be expanded in a complete set of
spherical harmonics $S_j^{m \; m'}$ transforming in the $(j/2,j/2)$
representation of $SU(2) \times SU(2) \equiv SO(4)$, where $j$ is any
non-negative integer, and $-j/2 \le m,m' \le j/2$. It is convenient to
denote the full set of indices $(j,m,m')$ by $\alpha$.  These obey an
orthonormality condition (we take the radius of the $S^3$ to be one)
\eqn\sortho{
\int_{S^3} S^\alpha S^\beta = \delta^{\alpha \bar{\beta}},
}
where $S^{\bar{\alpha}}$ denotes the complex conjugate of $S^\alpha$,
\eqn\scomcon{
(S_j^{m \; m'})^* = (-1)^{m+ m'} S_j^{-m \; -m'}.
}
The spherical harmonics are eigenfunctions of the Laplace operator on
the sphere,
\eqn\lapeq{
\nabla^2 S^\alpha = -j_\alpha (j_\alpha + 2) S^\alpha,
}
and under a parity operation transform with eigenvalue $(-1)^{j_\alpha}$.

A general vector field on the sphere may be expanded as a combination
of gradients of the scalar spherical harmonics plus a set of vector
spherical harmonics $\vec{V}_{j \pm}^{m \; m'}$.  These transform in
the $({j \pm 1 \over 2}, {j \mp 1 \over 2})$ representation of
$SO(4)$, where $j$ is a positive integer. Again, it is convenient to
denote the full set of indices $(j,m,m',\epsilon)$ by a single index
$\alpha$. These obey orthonormality relations
\eqn\vorth{\eqalign{
\int_{S^3} \vec{V}^\alpha \cdot \vec{V}^\beta &= 
\delta^{\alpha \bar{\beta}}, \cr
\int_{S^3} \vec{V}^\alpha \cdot \vec{\nabla} S^\beta &= 0 \; .
}}
Again $V^{\bar{\alpha}}$ indicates the complex conjugate of
$V^\alpha$, given by
\eqn\vcomcon{
({\vec V}_{j \pm}^{m \; m'})^* = (-1)^{m+ m'+1} 
{\vec V}_{j \pm}^{-m \; -m'} \; .
}
The vector spherical harmonics are eigenfunctions of parity with
eigenvalue $(-1)^{j+1}$, and satisfy
\eqn\lapcurl{\eqalign{
\nabla^2 \vec{V}^\alpha &= -(j_\alpha + 1)^2 \vec{V}^\alpha, \cr
\vec{\nabla} \times \vec{V}^\alpha &= 
-\epsilon_\alpha (j_\alpha + 1) \vec{V}^\alpha, \cr
\vec{\nabla} \cdot \vec{V}^\alpha &= 0.}}

Explicit expressions for the scalar and vector spherical harmonics may
be found in \Cutkosky.

\subsec{Spherical harmonic integrals}

The vertices of the mode-expanded Yang-Mills theory on $S^3$ have coefficients involving integrals over three spherical harmonics. We define
\eqn\sphconv{\eqalign{
B^{\alpha \beta \gamma} & \equiv \int_{S^3} S^\alpha S^\beta S^\gamma, \cr
C^{\alpha \beta \gamma} & \equiv 
\int_{S^3} S^\alpha \vec{V}^\beta \cdot \vec{\nabla} S^\gamma, \cr
D^{\alpha \beta \gamma} & \equiv \int_{S^3} \vec{V}^\alpha \cdot
\vec{V}^\beta S^\gamma, \cr 
E^{\alpha \beta \gamma} & \equiv \int_{S^3}
\vec{V}^\alpha \cdot (\vec{V}^\beta \times \vec{V}^\gamma) \; . \cr}}
It is also convenient to define
\eqn\moreconv{\eqalign{ \hat{B}^{\alpha \beta \gamma} & \equiv {1 \over
j_\alpha (j_\alpha + 2) j_\beta (j_\beta + 2)} \int_{S^3} (\vec{\nabla}
S^\alpha) \cdot (\vec{\nabla} S^\beta) S^\gamma \cr &={1 \over 2}
{(j_\alpha (j_\alpha + 2) + j_\beta (j_\beta + 2) - j_\gamma
(j_\gamma + 2)) \over j_\alpha (j_\alpha + 2) j_\beta (j_\beta + 2)}
B^{\alpha \beta \gamma}, \cr \hat{E}^{\a \b \g} & \equiv E^{\a \b \g}
(\e_\a (j_\a + 1) + \e_\b (j_\b + 1) + \e_\g (j_\g + 1)) \; . }}
These integrals were calculated in \Cutkosky, and the results may be
expressed as\foot{The expression for $C$ below differs by a factor
of two from the expression in \Cutkosky, but we believe that this
expression is correct.}
\eqn\threeints{\eqalign{ B^{\alpha \beta \gamma} &=
\left(\matrix{{j_\alpha \over 2} &{j_\beta \over 2} & {j_\gamma
\over 2} \cr m_\alpha&m_\beta&m_\gamma }\right)
\left(\matrix{{j_\alpha \over 2} &{j_\beta  \over 2} & {j_\gamma
\over 2} \cr m'_\alpha&m'_\beta&m'_\gamma}\right) R_1(j_\alpha,
j_\beta, j_\gamma), \cr C^{\alpha \beta \gamma} & =
\left(\matrix{{j_\alpha \over 2} &{j_\beta + \epsilon_\beta \over 2}
& {j_\gamma \over 2} \cr m_\alpha&m_\beta&m_\gamma }\right)
\left(\matrix{{j_\alpha \over 2} &{j_\beta - \epsilon_\beta \over 2}
& {j_\gamma \over 2} \cr m'_\alpha&m'_\beta&m'_\gamma}\right)
R_2(j_\alpha, j_\beta, j_\gamma), \cr D^{\alpha \beta \gamma} & =
\left(\matrix{{j_\alpha + \epsilon_\alpha \over 2} & {j_\beta +
\epsilon_\beta \over 2} & {j_\gamma \over 2} & \cr
m_\alpha&m_\beta&m_\gamma }\right) \left(\matrix{{j_\alpha -
\epsilon_\alpha \over 2} & {j_\beta - \epsilon_\beta \over 2} &
{j_\gamma \over 2} & \cr m'_\alpha&m'_\beta&m'_\gamma}\right) R_{3
\epsilon_\alpha \epsilon_\beta}(j_\alpha, j_\beta, j_\gamma), \cr
E^{\alpha \beta \gamma} & = \left(\matrix{{j_\alpha +
\epsilon_\alpha \over 2} & {j_\beta + \epsilon_\beta \over 2} &
{j_\gamma + \epsilon_\gamma \over 2} & \cr
m_\alpha&m_\beta&m_\gamma}\right) \left(\matrix{{j_\alpha -
\epsilon_\alpha \over 2} & {j_\beta - \epsilon_\beta \over 2} &
{j_\gamma - \epsilon_\gamma \over 2} & \cr
m'_\alpha&m'_\beta&m'_\gamma }\right) R_{4 \epsilon_\alpha
\epsilon_\beta \epsilon_\gamma}(j_\alpha, j_\beta, j_\gamma), \cr }}
where \eqn\rone{R_1(x,y,z)={ (-1)^{\sigma}\over\pi}\left(
\frac{(x+1)(y+1)(z+1)}{2} \right)^{1 \over 2}, }
\eqn\rtwoo{R_2(x,y,z)={
(-1)^{\sigma'}\over\pi}\left[\frac{(x+1)(z+1)(\sigma'-x)(\sigma'-y)(\sigma'-
z)(\sigma'+1)}{(y+1)}\right]^{1\over 2},}
\eqn\Rcombo{\eqalign{ R_{3 \e_x \e_z}(x,y,z) =& {(-1)^{\s +
(\e_x+\e_z)/2} \over \pi}\left({(y+1) \over 32
(x+1)(z+1)}\right)^{1 \over 2} \cr & \cdot \left( (\e_x(x+1) +
\e_z(z+1) + y+2)(\e_x(x+1) + \e_z(z+1)
 + y) \right. \cr
& \quad \left. (\e_x (x+1) + \e_z (z+1)  - y)(\e_x (x+1) +
\e_z(z+1) -
 y - 2) \right)^{1 \over 2}, \cr
R_{4 \e_x \e_y \e_z}(x,y,z) =& {(-1)^{\s'+1} \over \pi} {\rm
sign}(\e_x+\e_y+\e_z) \left({(\s'+1)(\s'-x)(\s'-y) (\s'-z) \over
4(x+1)(y+1)(z+1)}\right)^{1 \over 2} \cr &
\!\!\!\!\!\!\!\!\!\!\!\!\!\!\!\!\! \cdot \left( (\e_x(x+1) +
\e_y(y+1) + \e_z(z+1) +2 ) (\e_x(x+1) + \e_y(y+1) + \e_z (z+1)-2)
\right)^{1 \over 2}.
}}
Here, the right-hand sides of the equations are defined to be
non-zero only if the triangle inequality $|x-z| \le y \le x+z$
holds, and if $\s\equiv (x+y+z)/2$ (in $R_1$ and $R_3$) and $\s' \equiv
(x+y+z+1)/2$ (in $R_2$ and $R_4$) are integers. We also define
$R_{3+}\equiv R_{3++}=R_{3--}$, $R_{3-}\equiv R_{3+-}=R_{3-+}$,
$R_{4+}\equiv R_{4+++}$ and $R_{4-}\equiv R_{4++-}$.

\subsec{Identities for sums of spherical harmonics}

Using identities for 3j-symbols which may be found in \firstord, it
is straightforward to derive expressions for sums over $m$, $m'$,
and $\epsilon$ in various products of the spherical harmonic
integrals. For our calculations, we require:
\eqn\sphids{\eqalign{ \sum_{m's} B^{\a \bal \l} &= {1
\over \sqrt{2} \pi} (j_\alpha + 1)^2 \delta_{\lambda,0} , \cr
\sum_{m's, \epsilon} D^{\a \bal \l} &= {\sqrt{2} \over \pi}
\delta_{\lambda,0} j_\a (j_\a +2), \cr \sum_{m's, \epsilon} C^{\a
\d \g} C^{\bg \bd \bar{\a}} &= -2  R_2^2(j_\a, j_\d,
j_\g), \cr  \sum_{m's, \epsilon's} D^{\alpha \beta  \gamma}
D^{\bar{\alpha} \bar{\beta} \bar{\gamma}} &= 2
R^2_{3+}(j_\alpha,j_\gamma,j_\beta) + 2 R^2_{3-}(j_\alpha,
j_\gamma,j_\beta), \cr
 \sum_{m's, \epsilon's,  j_\gamma} D^{\alpha \beta \gamma}
D^{\bar{\alpha} \bar{\beta} \bar{\gamma}} &= {2 \over 3\pi^2}
j_\alpha (j_\alpha + 2) j_\beta (j_\beta +2), \cr \sum_{m's}
E^{\alpha \beta \gamma} E^{\bar{\alpha} \bar{\beta} \bar{\gamma}}
&= R^2_{4 \epsilon_\alpha \epsilon_\beta
\epsilon_\gamma}(j_\alpha, j_\beta, j_\gamma). }}

\appendix{C}{Useful formulae for dimensional regularization}

The following formulae are useful for our calculations in dimensional regularization of section 3.2:
\eqn\inteq{\int {d^d k \over (2 \pi)^d} {k^{2m} \over (k^2 + M^2)^n}
= {1 \over M^{2n-2m-d}} {1 \over (4 \pi)^{d \over 2} }
{\Gamma(n-m-{d \over 2}) \over \Gamma(n)}{\Gamma(m+{d \over 2})
\over \Gamma({d \over 2})},}
\eqn\forzeta{ \zeta(1+ \epsilon) = {1 \over \epsilon} + \gamma +
{\cal O}(\epsilon),}
\eqn\forgammaprime{\psi(z) \equiv {\Gamma'(z) \over \Gamma(z)},}
\eqn\forpsi{\psi(n) = - \gamma + \sum_{k=1}^{n-1}{1 \over k},}
\eqn\nforpsi{\psi({1 \over 2} + n) = - \gamma -2 \ln(2) + 2
\sum_{k=1}^{n}{1 \over 2k -1}.}

\appendix{D}{Regulating the Coulomb gauge}

For our calculations here and in \firstord, we have used the Coulomb
gauge, setting the divergence of the spatial gauge field to zero. In
this gauge, the time component of the gauge field and the ghosts
have no kinetic term, so their propagator contains a delta function.
For diagrams containing $A_0$ or ghost loops, we then get a
$\delta(0)$ factor. While this always cancels between the diagrams,
one may be concerned that certain finite terms have been missed.
Further, the singular propagators lead to ambiguities (which we did
not encounter here) in certain calculations when the time derivative
of the $A_i$ propagator must be evaluated at $t=0$.

To give an ambiguity-free definition of the Coulomb gauge, we can
replace the strict $\vec{\nabla} \cdot \vec{A} = 0$ condition by a
gauge-fixing action
\eqn\forsgf{S_{gf} = \int d^4 x \; \tr({1 \over 2 \xi} (\partial_0
A^0 + \xi \vec{\nabla} \cdot \vec{A} )^2),}
together with the corresponding ghost action
\eqn\forsgh{S_{gh} = \int d^4 x \; \tr({1 \over \xi} \partial_0
\bar{c} D_0 c +
\partial_i \bar{c} D_i c).}
The Coulomb gauge may be defined as the $\xi \to \infty$ limit of
this, but performing calculations at finite $\xi$ avoids any
$\delta(0)$ singularities (they show up as $\xi^{1 \over 2}$ terms
that cancel between the diagrams). In practice, we need only keep
the $\xi^{1 \over 2}$ and $\xi^{0}$ terms for each diagram. Using
this procedure, we can verify that no additional finite terms arise
in the cancellation of the $\delta(0)$'s in our calculation.

On the other hand, we have found that for certain other
calculations, the
naive Coulomb-gauge calculations can miss finite contributions. As
an example, in the two-loop Casimir energy on the sphere, a diagram
with two ${1 \over \xi} c A_0 c$ vertices gives a $\xi$-independent
contribution that would be missed if we set $\xi= \infty$ from the
start.

\listrefs

\end